\documentclass[prb,aps,twocolumn,showpacs,floatfix]{revtex4-1}
\usepackage[utf8x]{inputenc}
\usepackage{amsbsy}
\usepackage{geometry}
\usepackage{color}
\usepackage{graphicx}
\usepackage{color}
\usepackage{dcolumn}
\usepackage{amsmath}
\usepackage{amssymb}
\usepackage{amssymb}
\usepackage{amsfonts}
\usepackage[caption=false]{subfig}
\usepackage{xcolor}
\definecolor{darkgreen}{RGB}{0,160,48}

\usepackage[colorlinks=true, citecolor=black, linkcolor=black, 
urlcolor=black]{hyperref}

\newcommand{\beq}{\begin{equation}}
\newcommand{\eeq}{\end{equation}}
\newcommand{\beqn}{\begin{eqnarray}}
\newcommand{\eeqn}{\end{eqnarray}}
\def\qq{\mathbf{q}}
\def\GG{\mathbf{G}}
\def\kk{\mathbf{k}}

\begin{document}
\title{Breakdown of Herring's processes in cubic semiconductors for sub-terahertz longitudinal acoustic phonons}

\author{Maxime Markov}
\author{Jelena Sjakste}
\author{Nathalie Vast}
\affiliation{Laboratoire des Solides Irradi\'{e}s (LSI), \'Ecole Polytechnique - CEA/DRF/IRAMIS - CNRS UMR 7642,  91128 Palaiseau c\'edex, France}
\author{Romain Legrand}
\author{Bernard Perrin}
\affiliation{Institut de Nanosciences de Paris (INSP), Sorbonne Université, CNRS UMR 
7588, Case 840, 4 place Jussieu, 75252 Paris Cedex 05, France}
\author{Lorenzo Paulatto}
\email{lorenzo.paulatto@sorbonne-universite.fr}
\affiliation{Institut de Min\'eralogie, de Physique des Mat\'eriaux et de 
Cosmochimie (IMPMC), Sorbonne Université, CNRS UMR 
7590, IRD UMR 206, Case 115, 4 place Jussieu, 75252 Paris Cedex 05, France}

\begin{abstract}
In the present work we explain the anomalous behavior of the attenuation of the longitudinal acoustic phonon in GaAs as a function of the phonon energy $\omega$  in the sub-THz domain. 
These attenuations along the [100] direction show a plateau between 0.6 and 1~GHz at low temperatures. 
We found an excellent agreement between  measurements performed by some of us, and new \textit{ab initio} calculations of third-order anharmonic processes. The formation of the plateau is explained by the competition between different phonon-phonon scattering processes as Herring's mechanism, which dominates at low frequencies, saturates and disappears. The plateau is shown to be determined by the phononic final-state phase-space available at a given temperature. We predict that a change of scattering mechanism should also show up in the attenuation of silicon around 1.2-1.7~THz, and argue that the attenuation plateau is a general feature of cubic semiconductors.

\end{abstract}

\pacs{63.20.kg,63.20.dk,72.80.Cw,72.80.Ey,74.25.Ld,43.35.Cg}

\maketitle

\section{Introduction}\label{introduction}
Many efforts have been devoted to the measurement of the absorption of ultrasonic waves with frequencies below a few GHz\cite{Tucker:1972}. The results have been discussed in the framework of the Landau-Rumer or Akhiezer theories according to the frequency and temperature ranges in which they have been performed \cite{Bommel:1959,Mason:1964,Pomerantz:1965,Cottam:1974,Hasson:1975,Chavez:2014}.
On the other hand, high frequency phonon lifetimes are actively studied nowadays in link with thermal and thermoelectric transports, both experimentally \cite{Zeng:2015,Shi:2016}  and by computer simulations \cite{1.2822891,Mingo:2009,Luo:2013,Esfarjani:2011,Fugallo:2013,Markov:2016}.

However there are only very few studies about the damping of subterahertz acoustic waves although a full understanding of the attenuation of these waves is becoming crucial in several respects: \textit{i)} Efforts are devoted today to develop non-destructive methods for phonon imaging of deeply embedded nanostructures using picosecond acoustics, potentially very important for microelectronics \cite{Daly:2004}. It is thus important to know the absorption length of short acoustic pulses in standard semiconducting materials (silicon, germanium, sapphire, GaAs) commonly used for microelectronics. \textit{ii)} Quantum optomechanics can play a major role in quantum information: the challenge is to use resonators with higher frequencies (tens of GHz, instead of the current threshold of some hundreds of MHz), which would allow to observe the quantum regime at higher temperatures \cite{Fainstein:2013}. Then, a key problem is the limitation of the quality factor of such resonators due to the intrinsic phonon-phonon interaction intervening at higher temperatures \cite{Rozas:2009,Hamoumi:2018}.

Attenuation of subterahertz acoustic waves can be studied by the so-called “picosecond ultrasonic technique” \cite{Thomsen:1986}. In such experiments, either short acoustic pulses with a broad spectrum ranging from a few tens of GHz up to hundreds of GHz, or monochromatic coherent acoustic waves up to a few THz, can be generated and detected by metallic films, quantum wells or semiconducting superlattices \cite{Huynh:2011,Mante:2015,Huynh:2015}. The attenuation of these propagating sound waves can be measured over a large temperature range using samples with thickness going from a few micrometers up to millimetric size\cite{Hao:2001,Duquesne:2003,Daly:2009,Huynh:2015,Legrand:2016}. At low temperature in pure crystals, the damping of acoustic waves above 10~GHz is mainly due to three-phonon interactions, which corresponds to the so-called Landau-Rumer (LR) regime. Under such conditions, one can get access to accurate experimental information about lifetimes of individual longitudinal acoustic (LA) phonons.  At higher temperatures, 
the LR regime holds only for higher frequencies (above 100~GHz).  At lower frequencies, the three-phonon perturbation description breaks down and the system crosses into the Akhiezer regime, where
the collective response of phonon populations under the effect of the strain field induced by the acoustic wave has to be taken into account. In the intermediate regime between the LR and Akhiezer ones, the finite lifetime of phonons interacting with the exciting longitudinal acoustic wave should also be taken into account.

In the LR regime, the interaction of three phonons is the main mechanism of the phonon decay~\cite{Maris:1971}. 
For a longitudinal acoustic wave propagating in an anisotropic crystal, Herring  has pointed out a dominant three-phonon coalescence mechanism involving the scattering of the excited longitudinal acoustic wave by a slow transverse phonon into a fast transverse phonon: $\omega_{LA}+\omega_{TA_s} \rightarrow \omega_{TA_f}$, where
$\omega$ are the phonon frequencies, $LA$ refers to the longitudinal branch, and $TA_s$, $TA_f$ to the transverse branches with respectively the lowest and highest sound velocities~\cite{Herring:1954a}.  Herring's processes are dominant because they allow the coupling of low energy longitudinal phonons close to the $\Gamma$ point, to transverse phonons which have a much higher wavevector $\mathbf{q}$ and, thus, a large density of states. Using symmetry arguments,  Herring predicted  that the contribution of this process should be proportional to $\omega^2T^3$ in cubic crystals, where $T$ is the temperature. However, this dependence only holds in a very limited frequency and temperature domain~\cite{Legrand:2014} and no general theoretical statement has been done beyond these limits yet. Recent experimental measurements on the absorption of subterahertz longitudinal waves made by some of us in gallium arsenide \cite{Legrand:2016} showed that after a steep increase, the attenuation exhibits an unexpected plateau as a function of the excitation frequency in the 700~GHz~-~1~THz range. This plateau has been ascribed, on the basis of strong hypotheses, to a breakdown of the Herring processes~\cite{Legrand:2016}.

In this paper, we consider acoustic waves of frequency  up to~1 or 2 THz, at temperatures between 50 and 300 K for silicon, and down to 2~K for gallium arsenide. We study acoustic phonon attenuation in GaAs and Si by means of \emph{ab initio} calculations based on the density functional perturbation theory \cite{Paulatto:2013}.
We show that the attenuation plateau observed experimentally~\cite{Legrand:2016} in GaAs can be fully reproduced by \emph{ab initio} calculations, and can be explained
by the density of final states available for different phonon-phonon scattering processes. More precisely, the plateau is explained by the 
rapid decrease of the probability of Herring's processes in the 700~GHz~-~1~THz range, whereas the probability of other phonon-phonon scattering processes
is found to increase. Moreover, we predict that a similar  plateau in the attenuation of acoustic phonons should be observed in silicon around 1.2~-~1.7~THz.
 
The paper is organized as follows: first, we  present our \emph{ab initio}
method to compute phonon attenuation due to three-phonon interaction, and provide technical details which concern in particular the convergence of phonon lifetimes.
Secondly, we present our calculations for the attenuation of acoustic phonons in GaAs, comparing   with experimental data. We discuss the origin of the attenuation plateau found both theoretically and experimentally and examine the roles of the joint density of final states available for different phonon-phonon scattering processes, in particular for Herring's processes. The role of the matrix elements of the phonon-phonon interaction is investigated. 
In the fourth and  fifth sections, we present our predictions for the attenuation in silicon, we justify our conclusion that Herring's breakdown is general to cubic semiconductors, and discuss the conditions of its observation.
Finally, we discuss the applicability of the long wavelength approximation (LWA) for the anharmonic coupling coefficients which we express in terms of third order elastic constants \cite{Simons:1957,Tamura:1985,Berke:1988}.

\section{Method}\label{cmethod}

The computational method to obtain fully \textit{ab initio} the matrix
elements of the phonon-phonon interaction has been described in detail in reference~\onlinecite{Paulatto:2013}, and in papers 
cited therein. For the sake of completeness, we review the theory involved using the compact notation of reference~\onlinecite{Fugallo:2013}. 
To first order in perturbation theory, the intrinsic phonon-phonon interaction is described by one  single Feynmann's diagram (the ``bubble'' diagram)\cite{Calandra:2007a}. This diagram describes both the decay (fission) process, where the initial phonon  decays into two phonons, and the coalescence (fusion) process, where the phonon is scattered by another one to create a new phonon. In both kinds of processes, the total energy and the crystal momentum have to be conserved, the latter modulo a vector $\GG$ of the reciprocal lattice because of periodic boundary conditions. 

Identifying the phonon by its wavevector $\qq$ and branch index $j$, and referring to the  phonon energy $\omega_{\qq,j}$ and Bose-Einstein occupation  $n_{\qq,j}$, we define the inverse of the phonon lifetime $\tau_{\qq,j}^{-1}$ as a sum in the reciprocal space\cite{Lazzeri:2003b,Calandra:2007a}:

\begin{small}
\begin{align}\label{phph}
 &\left(\tau \right)^{-1}_{\qq,j}  = 
 \frac{\pi}{\hbar N_0} \!\sum_{\qq',j',j''} 
 \left| V^{(3)}(\qq j,\qq' j', \qq'' j'') \right|^2 \times \nonumber\\
 &\Big[ 
  (1+n_{\qq',j'} + n_{\qq'',j''}) \delta(\hbar\omega_{\qq,j}-\hbar\omega_{\qq',j'}-\hbar\omega_{\qq'',j''})
 \nonumber\\
 &\phantom{\Big[}+  
 2(n_{\qq',j'} - n_{\qq'',j''}) \delta(\hbar\omega_{\qq,j}+\hbar\omega_{\qq',j'}-\hbar\omega_{\qq'',j''}) 
 \Big]\,.
\end{align}
\end{small}

In this equation, we have introduced $V^{(3)}(\qq j,\qq' j', \qq'' j'')$, the matrix elements of the phonon-phonon interaction, that are closely related to the third-order derivatives of the total electronic energy, calculated with respect to three phonons. We compute them fully \textit{ab initio}, using density functional perturbation theory\cite{Baroni:2001} and the ``$2n+1$'' theorem.\cite{Debernardi:1995,Lazzeri:1998,Lazzeri:2002} The conservation of momentum is imposed by requiring that  $\qq'' = -(\qq+\qq')$. In our case, $\qq$ is the
momentum of the initial acoustic wave (phonon) for which the attenuation is calculated, $\qq'$ runs on a regular grid of  $N_0$ points in the whole Brillouin zone, 
and $\qq''$ is fixed by momentum conservation. 

The first term of eq.~\ref{phph} describes decay processes,
while the second one describes coalescence processes. Because of the magnitude of the Bose-Einstein occupation terms, decay processes usually dominate. However the conservation of energy  (expressed by the first Dirac $\delta$-function
in eq.~\ref{phph}) requires that the two created phonons $(\qq',j')$ and $(\qq'',j'')$ have values of the energy lower than that of the initial phonon. As a consequence, for low energy initial acoustic phonons, there tends to be no suitable final states for decay, so that the contribution of decay terms to the sum of eq.~\ref{phph} is negligible. On the other hand, the second term of eq.~\ref{phph} describes  coalescence processes whose contribution to the sum of eq.~\ref{phph} is small for optical phonons, but turns out to dominate for low-energy acoustic phonons, as predicted by Herring, but with limitations that are the subject of the following sections. 

The attenuation of ultra-sound waves is related to the acoustic field amplitude by the following relation:
\begin{align}\label{attenuation}
 A_z(T) = A_0 e^{-\alpha(T) z}, 
\end{align}
where $A_z$ is the measured amplitude, $A_0$ is the amplitude of the generated pulse,  $z$ is the distance from the pulse source and $\alpha$ is the attenuation.
The attenuation is inversely proportional to the phonon lifetime and its group velocity \textit{i.e.},  to the derivative of the phonon frequency with respect to the wavevector,\footnote{This derivative has to be done with some care, to avoid problems when phonon branches cross or are degenerate. Details can be found in section IV of reference~\onlinecite{Fugallo:2013}.} $\mathbf{v}_{\qq,j} = \nabla_\qq \omega_{\qq,j}$. The attenuation reads: 
\begin{align} \label{computed_attenuation}
\alpha_{\qq,j}  =& \frac{1}{2 |\mathbf{v}_{\qq,j}|\, \tau_{\qq,j}}\,.
\end{align}
In order to understand which decay mechanisms contribute to the attenuation, we will study the temperature-dependent joint phonon density of states (T-JDOS) which gives  information about the phase space available for scattering processes involving a specific phonon ({\bf q},~j). In its most general form the definition of the T-JDOS, $\theta_{\qq,j}$, is based on equation~\ref{phph}, with matrix elements set to a constant value :

\begin{small}
\begin{align}\label{ljdos}
 &\theta_{\qq,j}  = 
 \frac{\pi}{\hbar N_0} \!\sum_{\qq',j',j''}  \times \\
 &\Big[ 
  (1+n_{\qq',j'} + n_{\qq'',j''}) \delta(\hbar\omega_{\qq,j}-\hbar\omega_{\qq',j'}-\hbar\omega_{\qq'',j''})
 \nonumber\\
 &\phantom{\Big[}+  
 2(n_{\qq',j'} - n_{\qq'',j''}) \delta(\hbar\omega_{\qq,j}+\hbar\omega_{\qq',j'}-\hbar\omega_{\qq'',j''}) 
 \Big]\,\nonumber.
\end{align}
\end{small}

If we only take the first line of the definition, we get the decay-specific T-JDOS ($\theta^D_{\qq,j}$), while the second line would give the coalescence part ($\theta^C_{\qq,j}$). Furthermore, we can restrict the latter to  processes that are described by Herring's mechanism: coalescence processs where a LA phonon combines with a slow TA phonon to form a fast TA.
In practice, this can be done by imposing that all of the phonons involved are acoustic ones, that ($\qq,j$) is longitudinal, 
and that $\omega_{\qq',j'} \leq \omega_{\qq'',j''}$. With these constraints, we get the Herring contribution to the T-JDOS ($\theta^H_{\qq,j}$). We will see in section \ref{sec:origin-gaas} up to which point the Herring T-JDOS is a good approximation for the full decay mechanism.

\section{Computational method}\label{cdetails}

\subsection{Calculation details}

Calculations were performed with the \textsc{Quantum-ESPRESSO} code \cite{Giannozzi:2009,Giannozzi:2017} using the LDA-PZ\cite{Perdew:1981} exchange and correlation functional. This functional is less popular than the PBE\cite{Perdew:1996} one.  It gives however a theoretical lattice parameter close to the experimental one in cubic semiconductors, which is extremely important to obtain a good description of the phonon dispersion.
For silicon, we have used pseudopotentials from the Fritz-Haber-Institut library\cite{Fhipp}. The pseudopotentials for gallium and arsenic were  generated
with the FHIpp code \cite{Fhipp}, and are the same ones as those used in references \onlinecite{Botti:2002,Botti:2004,Sjakste:2006}.  

We constructed a crystal unit cell and relaxed it to the theoretical lattice parameter, which was found to be 10.591~Bohrs for GaAs and 10.167~Bohrs for Si. The use of the experimental lattice parameters for both GaAs and Si was found to yield very little difference in the results. The Kohn-Sham set of equations was integrated in the reciprocal space over a shifted Monkhorst-Pack\cite{Monkhorst:1976} grid of $12\times 12\times 12$ (GaAs) or $8 \times 8 \times 8$ (Si) $\kk$-points. The use of a grid shifted from the $\Gamma$ point was crucial to obtain properly converged results in GaAs, especially in the calculation of effective charges from the linear response theory. We used a kinetic  cutoff energy of 45~Ry, for which phonon frequencies at the $\Gamma$ point were converged within one cm$^{-1}$. For the present application, we decided to converge very strictly all parameters, as the cost of the \textit{ab initio} calculation using the ``$2n+1$'' theorem is negligible with respect to the integration over the Brillouin zone in equation \ref{phph}.

\subsection{Convergence of the inverse lifetime}

Indeed, equation \ref{phph} includes two Dirac delta functions that enforce the conservation of energy. To make the integration feasible, these delta distributions are replaced by Gaussian functions of finite width, under the condition that the chosen width is smaller than energy differences of phonons involved in the process. In order to obtain good quality results, the sum over  $\qq$ points of the reciprocal space must be done over a grid which is sufficiently fine to sample a significant number of non-zero scattering processes. 
However, the number of points increases rapidly as we tackle low temperature and low energy phonons, because the finite width of the Gaussian has to be reduced to maintain the computational accuracy.

Such a fine grid cannot be directly computed from density functional perturbation theory. We thus exploit the fact that phonon dynamical matrices and  phonon-phonon coupling matrices change smoothly in reciprocal space, and use Fourier transform to pass from a relatively coarse grid of points, computed directly \textit{ab initio}, to short-ranged two- and three-body force constants in real space. The latter quantities are interpolated in real space, assuming that they are short ranged. A second Fourier transform, back to the reciprocal space, allows to compute  phonon dynamical matrices and  phonon-phonon coupling matrices at any $\qq$ on a ultrafine grid. Details of the interpolation scheme can be found in Ref.~\onlinecite{Paulatto:2013}.

In the present application, we have computed the phonon dynamical matrices \textit{ab initio} on an initial $8 \times 8 \times 8$ $\qq$-point grid in the reciprocal space, and the phonon-phonon coupling matrices on a $4 \times 4 \times 4$ $\qq$-point grid. 
To evaluate the integral of equation~\ref{phph} the code does not employ symmetry operations to reduce the grid size, hence an uniform $\Gamma$-centered grid is not the best choice for the integration, as a large number of equivalent points would be included in the integral. We found instead that the use of a randomly shifted grid, which is incommensurate to any symmetry operation, yields the fastest convergence. Although symmetry is formally broken, we checked that it is numerically recovered at convergence. All of the phonon lifetime calculations have been performed with a grid containing as many as $199 \times 199 \times 199$ points. The finite smearing was chosen to be 0.2~cm$^{-1}$ for GaAs and 0.5~cm$^{-1}$ for Si, respectively. Reducing the grid size to $39 \times 39 \times 39$  $\qq$-point  was proven to be sufficient  in  almost every case, when used with a large smearing value of 2~cm$^{-1}$ (66~GHz). It was however insufficient to reproduce the correct behavior of the attenuation when the initial-phonon frequency was smaller than  100~GHz.

\section{Results for Gallium Arsenide}

\subsection{Attenuation of LA phonons along [100]}
\begin{figure*}
  \subfloat[a][ 50~K - $\omega$ dependency]{\includegraphics[width=0.48\textwidth]{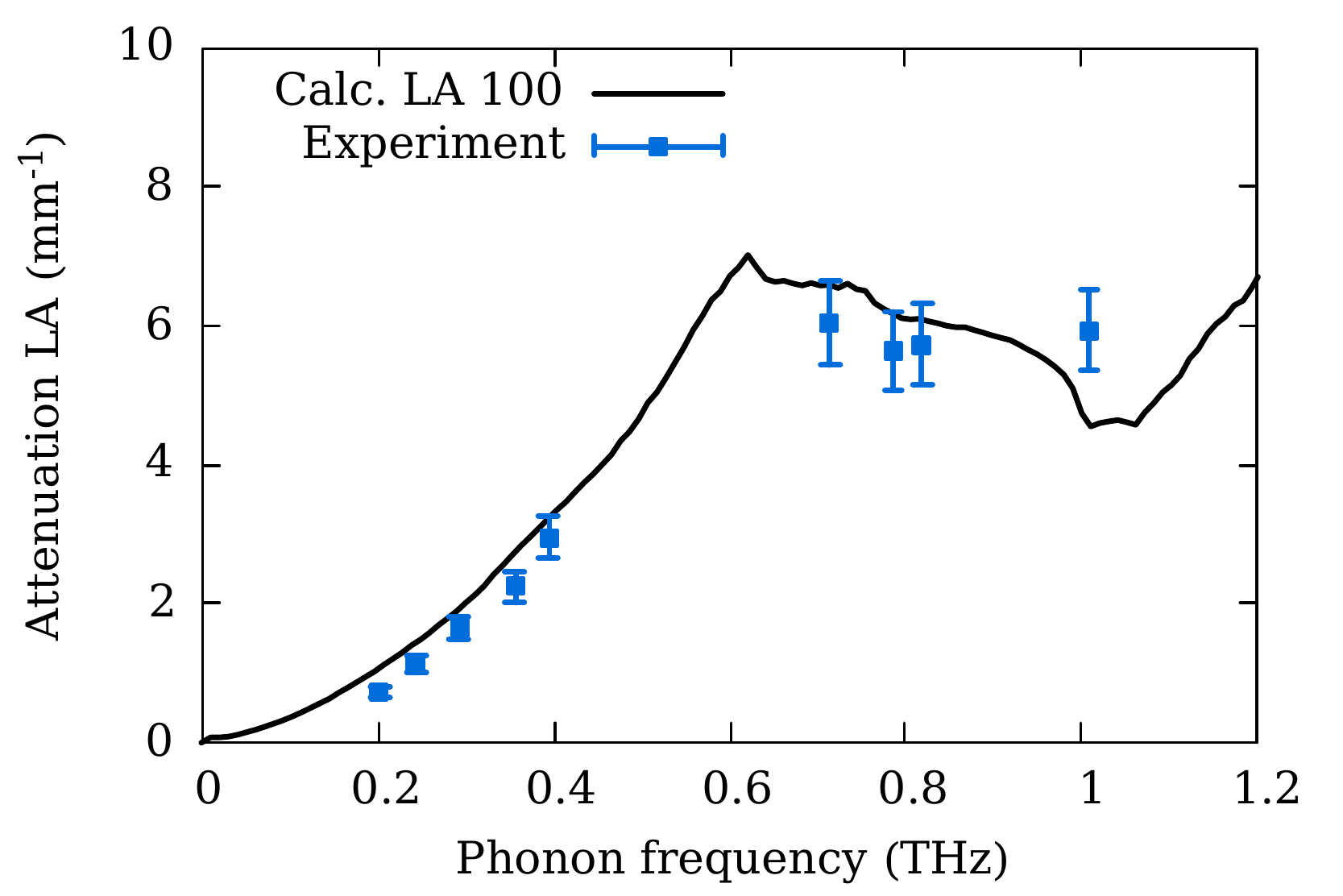}}
  \subfloat[b][ 713~MHz - $T$ dependency]{\includegraphics[width=0.48\textwidth]{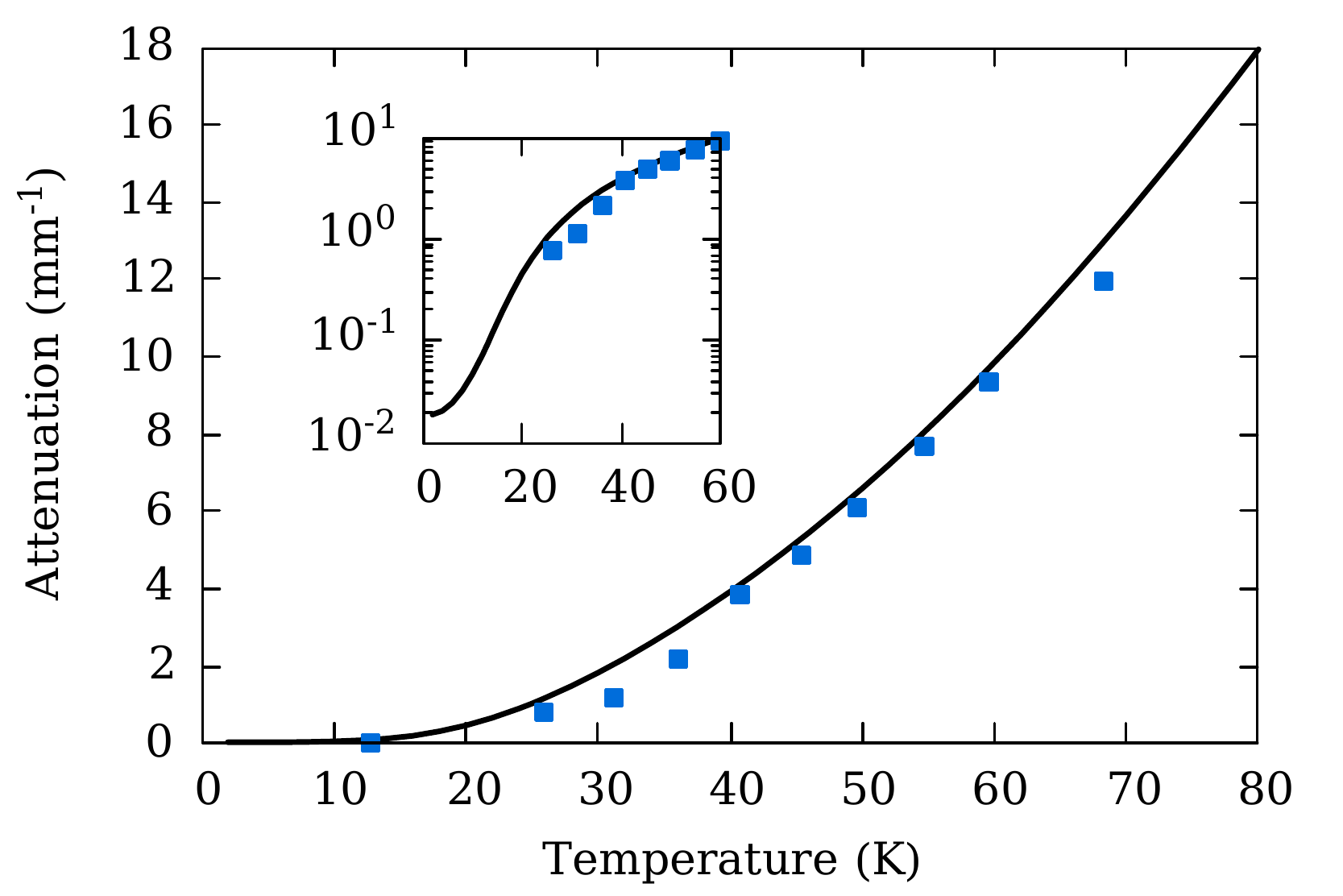}}
\caption{Phonon attenuation in GaAs: comparison of calculations with experiments (a) At fixed temperature (50~K) as a function of phonon frequency along the $[100]$ direction. (b) At fixed energy ($\omega=713$~GHz) as a function of temperature. The inset shows data on a logarithmic scale for $\alpha$ to magnify the low-temperature behavior. Similar pictures for all of the points measured in Ref.~\onlinecite{Legrand:2016} are provided as supplementary materials.}
  \label{fig:expcomp}
\end{figure*}
We  focus on the case for which we are able to directly compare experimental results and calculations: the longitudinal phonon branch along the $[100]$ direction. As shown in figure~\ref{fig:expcomp}a, the agreement is remarkable. Not only does the calculation correctly reproduce the qualitative behavior of $\alpha(\omega)$, and the presence of the plateau between 600~GHz and 1~THz, but also the experimental and calculated absolute values of the attenuation are very close. We stress that no renormalization has been applied to the data: the values of the attenuation directly come from the \textit{ab initio} calculations.   

In figure~\ref{fig:expcomp}b, we examine the behavior of $\alpha$ at the fixed frequency value of 713~GHz, as a function of the temperature, and compare it with the experimental data. The low-temperature region is magnified in the inset of figure 2. The agreement between the \emph{ab initio} calculations and the experimental data (starting from 20~K) is strikingly good even at very low temperature. We  report as supplementary materials the comparison for eight different frequencies, with similar conclusions.

At this point it is important to note that the experimental data correspond only to the relative change of attenuation with temperature defined by:

\begin{equation}
\alpha(T)-\alpha(10~K)=-\frac{1}{z}\ln(\frac{A_z(T)}{A_z(10~K)}),
\end{equation}
where $A_z$ is defined in eq.~\ref{attenuation}, while the \textit{ab initio} data is the absolute value of the attenuation given by equation \ref{computed_attenuation}.  

Thus, the experimental points would eventually go exactly to zero, at zero temperature, while the theoretical data, which take into account spontaneous decay processes, have therefore a finite limit at 0~K. The extremely good agreement above approximately 30~K implies that, indeed, the experimental points can also be considered as absolute values of the sound attenuation. The  behavior of the 
\emph{ab initio} data displayed in figure~\ref{fig:expcomp}b is typical of what could be expected from the temperature dependence of the inverse lifetime: a slow growth starting from a finite value at low temperature, which eventually becomes linear, in this case above 90~K (not shown). 

\begin{figure*}
  \subfloat[a][ 50~K]{\includegraphics[width=0.48\textwidth]{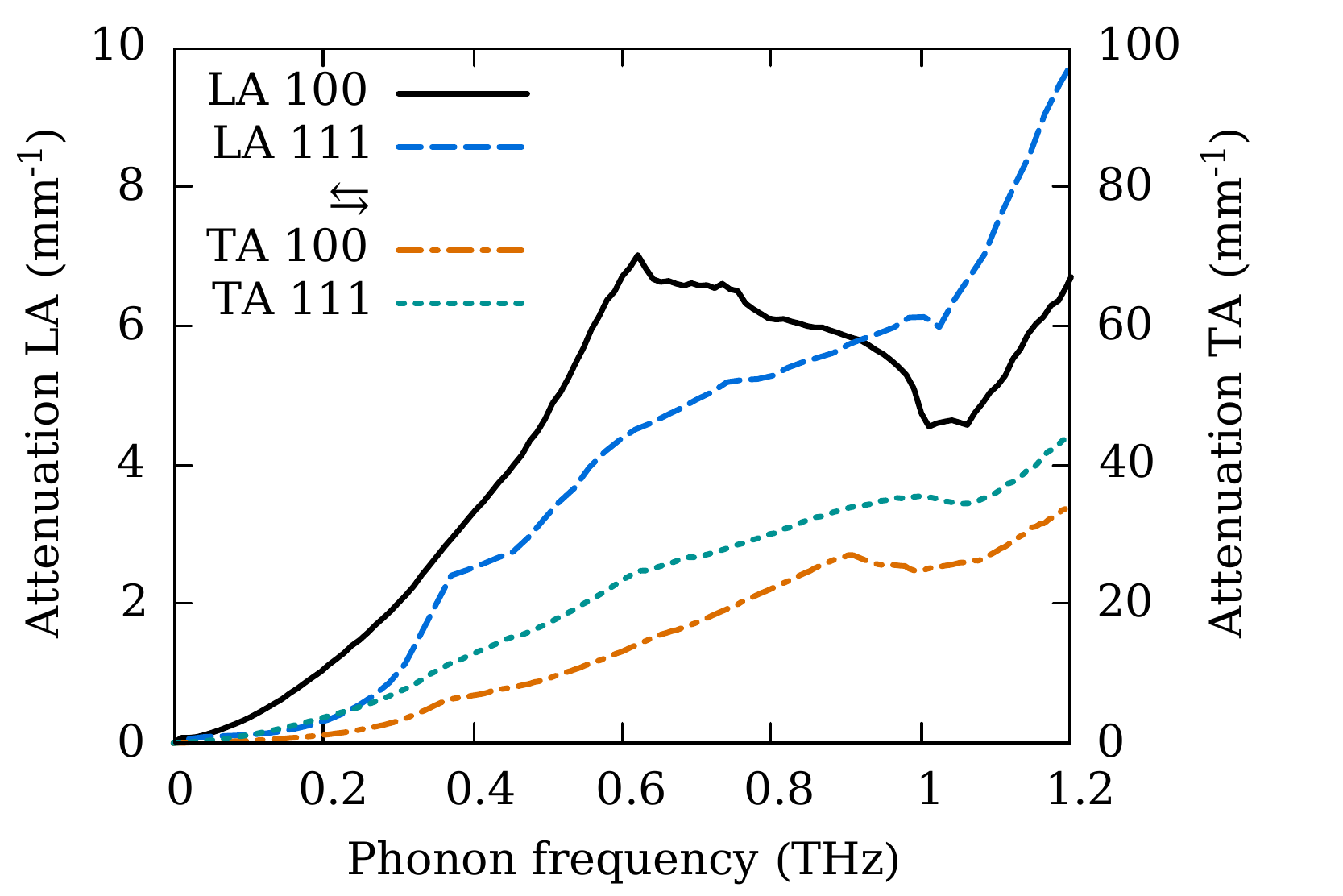}}
  \subfloat[b][ 300~K]{\includegraphics[width=0.48\textwidth]{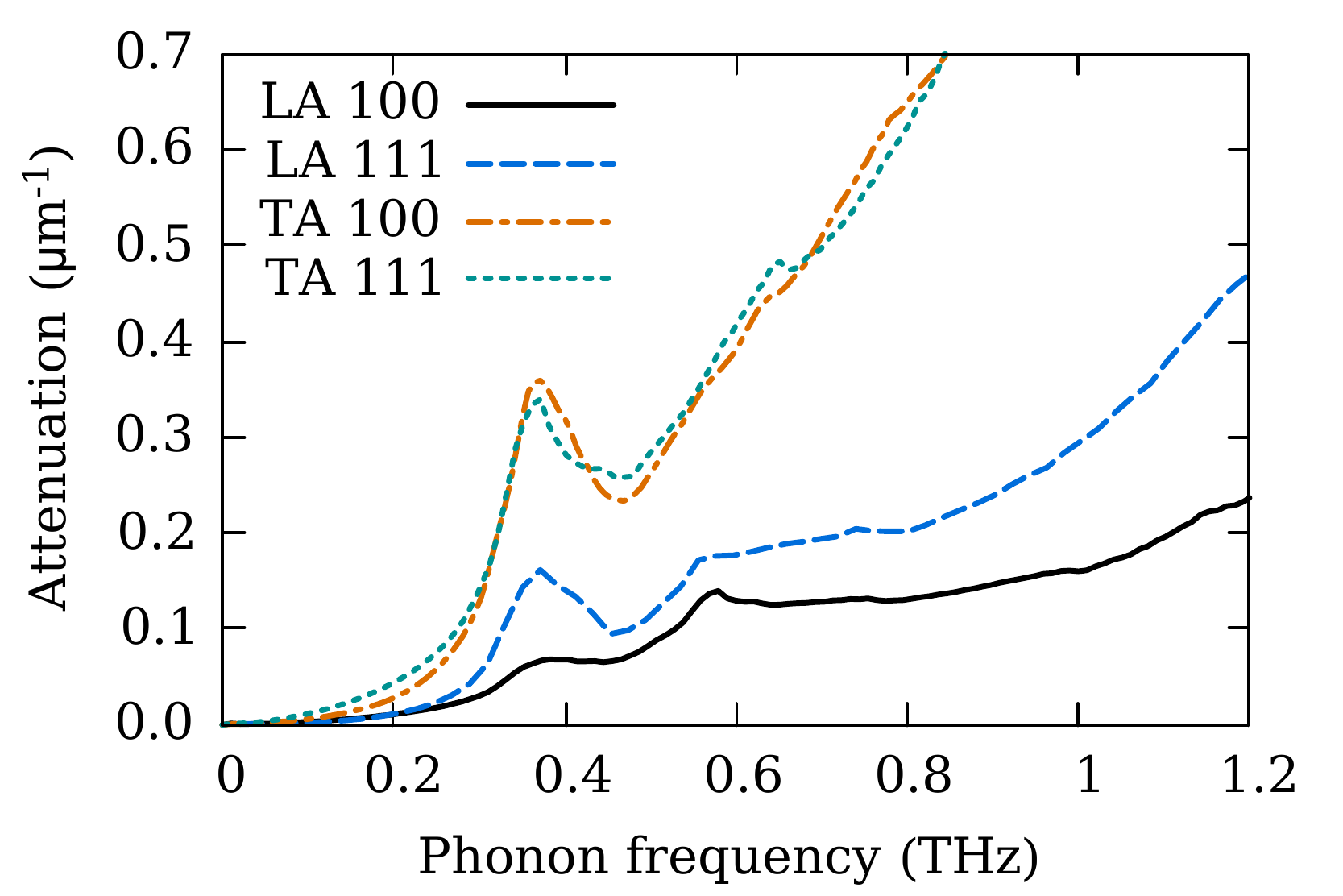}}
  \caption{GaAs. Attenuation of longitudinal and transverse phonons along the $[100]$ and $[111]$ directions at 50~K (left panel) and 300~K (right panel). Note the change of scale between the two temperatures (panels), as well as the different axis for the TA mode (on the right) in the left panel.}
  \label{fig:ofomega}
\end{figure*}

\subsection{Comparison of LA and TA modes and propagation directions}

In figure \ref{fig:ofomega} we extend the computation to the [111] direction, and to phonon modes which are not readily accessible by experiment. 

The behavior of the computed attenuation for transverse phonons is different from that of the longitudinal one: while for the LA branch we have a very clear plateau between 600~GHz and 1~THz, for the TA branches the attenuation only shows a shoulder around 1~THz.

We also report the attenuation of LA and TA phonons along the $[111]$ direction. None of the attenuations of the two branches has such a clear plateau as the LA phonon attenuation along the $[100]$ direction, but we observe a change of behavior around 1~THz, where the attenuation starts to grow faster as the phonon energy increases. 

At room temperature (panel~b, figure~\ref{fig:ofomega}), the plateau observed for the longitudinal wave along the [100] direction moves to lower energies, around 400~GHz, and extends over a frequency range smaller than the one at 50~K. Finally, a small dip is predicted at 400~GHz for the longitudinal wave along the [111] direction as well as for the transverse waves in both $[100]$ and $[111]$ directions.

\subsection{Results at cryogenic temperatures}

Some  measurements of the phonon attenuation in GaAs, additional to those of ref.~\cite{Legrand:2016}, are available in ref.~\onlinecite{Kent:2002}. The mean-free path (MFP) of the 650~GHz LA phonon along $[111]$ was measured at a temperature of 2~K, the reported value being 0.8~mm. At such a low temperature, the intrinsic phonon-phonon scattering is small with respect to the scattering of phonon with isotopic disorder and potentially with other lattice defects and impurities. We have simulated this MFP including only intrinsic scattering (not shown), and found a value of 40~mm. However when we included isotopic scattering, as done in reference~\onlinecite{Fugallo:2013}, the MFP was reduced to 2.7~mm, which we consider as an acceptable estimation of the intrinsic MFP.  Indeed, the difference between our result and the experiment of ref.~\onlinecite{Kent:2002} is the signature of the presence of additional scattering sources in the experiment, \textit{i.e.} lattice defects and surface roughness.

\subsection{Discussion on the frequency dependence of the attenuation}

To understand the origin of the plateau, we study whether it comes from the 
behavior of  transition matrix elements, or whether it is a property of the phonon dispersion, modulated by the matrix elements.
\subsubsection{Temperature-dependent joint-density of states\label{sec:origin-gaas}}

\begin{figure}
  \includegraphics[width=0.48\textwidth]{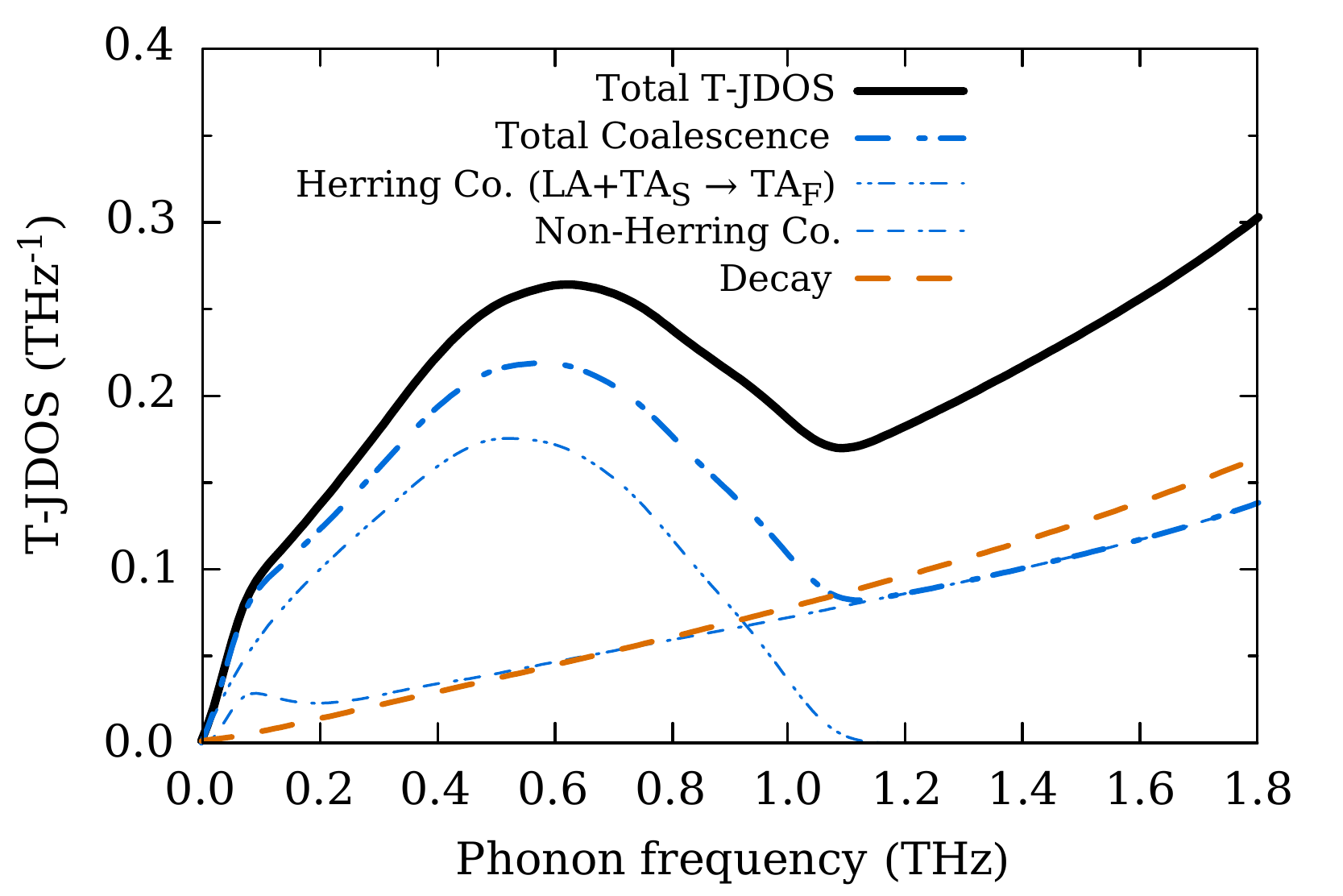}
\caption{Temperature-dependent joint-density of states (T-JDOS) at 50~K for the LA branch of GaAs along the $[100]$ direction. The total T-JDOS (thick black solid line) is decomposed into  decay (dashed orange) and coalescence (dot-dashed blue) processes, the latter being further divided into Herring's (LA$\rightarrow$TA$_S+$TA$_F$, blue dash-dotted lines) and Non-Herring's (blue long-dashed lines) processes.}
\label{fig:ljdos-gaas}
\end{figure}

 
We have computed the temperature-dependent joint-density of states (T-JDOS) of eq. \ref{ljdos} in the sub-THz region of the acoustic phonons in  GaAs, along the $[100]$ and $[111]$ high symmetry directions at two temperatures, 50~K and 300~K, as well as contributions
to T-JDOS from different scattering channels. We have observed that the T-JDOS is almost perfectly isotropic at the energies under consideration; furthermore the effect of temperature is limited to a change in magnitude, both absolute and between different contributions. 
For these reasons, in figure~\ref{fig:ljdos-gaas}, we only report the T-JDOS at 50~K and along the $[100]$ direction, which we discuss below. 

As one can see from figure~\ref{fig:ljdos-gaas}, for initial phonon frequencies under 1~THz, 
it is indeed the Herring mechanism that dominates over other phonon-phonon scattering mechanisms.  
However, while the contribution of other decay and coalescence events grows linearly when the frequency increases,  the Herring contribution
 saturates around 500~GHz. At this frequency, Herring's mechanism is still contributing by more than one half, up to two thirds, to the total value of the  T-JDOS, but it quickly looses importance until it completely dies off just beyond 1~THz. 
This breakdown of Herring's processes produces the change of slope in the total T-JDOS, which is the origin of the attenuation plateau experimentally observed in GaAs at low temperatures (fig.~\ref{fig:ofomega} and Ref.~\onlinecite{Legrand:2016}).

At 300~K (results not shown in the main text)\footnote{see supplementary materials for detailed figures for both materials, several temperatures and several high symmetry directions.}, the global picture is found to be very similar, but the steepness of the non-Herring contributions is higher than at 50~K.  This moves the maximum of the total T-JDOS down towards lower energies, as Herring's processes are drowned sooner by the other scattering mechanisms than at 50~K. Because the Herring contribution picks up very quickly from zero, it may appear as a short sharp peak or only as a change in the steepness of the total T-JDOS: the exact effect will depend crucially on the details of the matrix elements which govern the relative magnitude of the different mechanisms.

\subsubsection{Final states for phonon scattering}

On the one hand, T-JDOS plays the major role and 
it is the superimposition of the hill-shaped curve of the Herring mechanism and of the linear curve of the other processes, that results in the formation of the plateau. On the other hand, the matrix elements are still important: they fix the relative magnitude of scattering mechanisms and enable (prevent) the transitions which are allowed (forbidden) by symmetry, giving a much more complex shape and anisotropy of the attenuation (figures~\ref{fig:ofomega}) with respect to the simple and isotropic shape of the T-JDOS.

\begin{figure}
  \subfloat[a][150~GHz, {$[100]$}]{\includegraphics[width=0.245\textwidth]{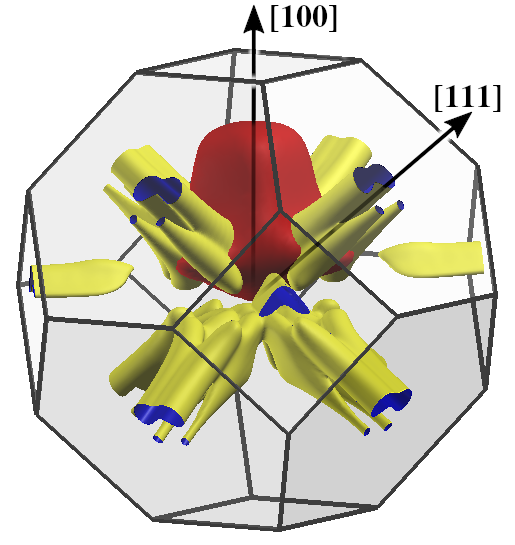}}
  \subfloat[b][900~GHz, {$[100]$}]{\includegraphics[width=0.245\textwidth]{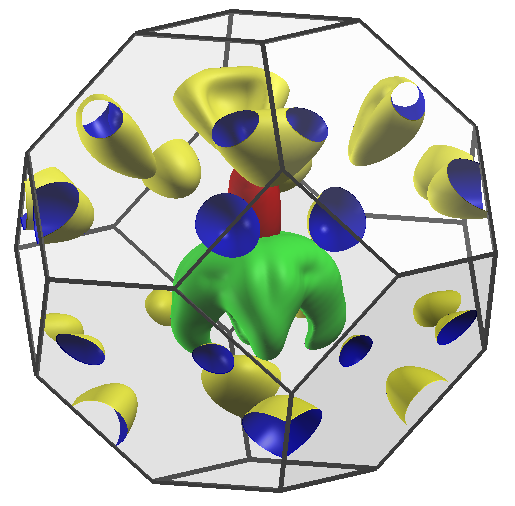}}\\
  \subfloat[c][150~GHz, {$[111]$}]{\includegraphics[width=0.245\textwidth]{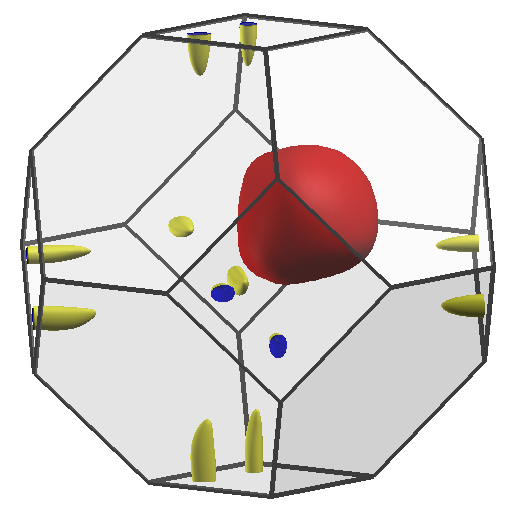}}
  \subfloat[d][900~GHz, {$[111]$}]{\includegraphics[width=0.245\textwidth]{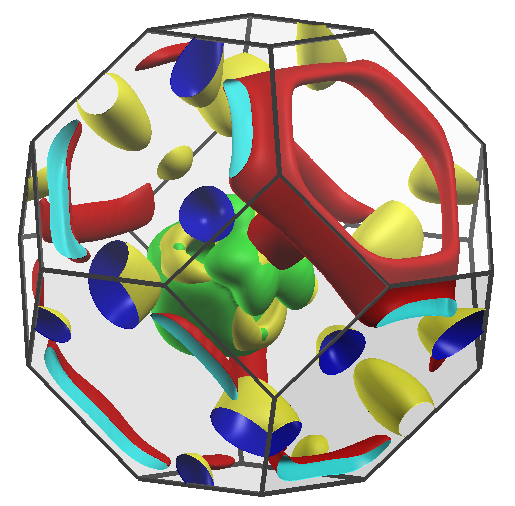}}
  \caption{GaAs at 50~K. Brillouin volume spanned by the wavevectors of the phonons involved in the scattering of LA phonons along the directions $[100]$ (panels a, b) and $[111]$ (panels c, d). The initial state is either at 150~GHz (panels a, c) or at 900~GHz (panels b, d). The volume inside the surfaces contributes 75\% to the scattering events. Surface-outside/inside colors allow to identify the final-state phonon  in the following way: TA slow is yellow/blue, TA fast is red/cyan,  and LA is green/magenta. The $[100]$ direction is oriented upward.}
  \label{fig:fstate-gaas}
\end{figure}

Furthermore, the scattering processes contributing to the attenuation strongly depend  on the direction because of the selection rules embedded in the matrix elements. In figure~\ref{fig:fstate-gaas} we have decomposed the contribution to the attenuation by convoluting equation~\ref{phph} with $\delta(\qq'-\bar\qq)$:
\begin{small}
\begin{align}\label{finalphph}
 &\left(\tau^{\qq,j,j'}(\bar\qq)\right)^{-1}  = \\ 
 &\frac{\pi}{\hbar N_0} \!\sum_{\qq',j''} \delta(\qq'-\bar\qq)
 \left| V^{(3)}(\qq j,\qq' j', \qq'' j'') \right|^2 \times \nonumber\\
 &\Big[ 
  (1+n_{\qq',j'} + n_{\qq'',j''}) \delta(\hbar\omega_{\qq,j}-\hbar\omega_{\qq',j'}-\hbar\omega_{\qq'',j''})
 \nonumber\\
 &\phantom{\Big[}+  
 2(n_{\qq',j'} - n_{\qq'',j''}) \delta(\hbar\omega_{\qq,j}+\hbar\omega_{\qq',j'}-\hbar\omega_{\qq'',j''}) 
 \Big]\nonumber\,.
\end{align}
\end{small}
We plot the resulting  $(\tau^{\qq,j,j'}(\bar\qq))^{-1}$
inside the Brillouin zone, with different color depending on $j'$. We have examined this decomposition for GaAs at 50~K, to study the attenuation of the LA phonons at 150~GHz, where the Herring processes dominate the scattering, and at 900~GHz, where Herring's processes have become minority. The iso-surfaces are color-coded depending on the band number of the final state. 

In particular, the yellow and red surfaces contain the $\hbar\qq$ vectors responsible for Herring's scattering. We can see how at low energy, along $[100]$ (fig.~\ref{fig:fstate-gaas}.a) the picture is consistent with the Herring model, with most of the scattering events occurring along the $[100]$ direction, the red volume, with also an important amount of scattering around the $[111]$ directions (the yellow volumes).
A significant fraction of the BZ (around 14\% of the BZ volume) is actively participating in the scattering. However, at the same energy but along the $[111]$ direction (panel (c)), only a smaller volume around $[111]$ and some small pockets close to the surface of the BZ in direction $[100]$ are Herring-active (about 4\% of the BZ volume). One sees in figure~\ref{fig:ofomega} by how much the attenuation at 150~GHz is  smaller along $[111]$ than along $[100]$. In panels b and d of figure~\ref{fig:fstate-gaas}, we can see that as we move towards higher energies of the initial phonon, the scattering mechanisms become more complex, and they involve all the bands (including the longitudinal band, in green) and entail large chunks of the BZ.

\subsubsection{Average matrix element}
\begin{figure}
  \begin{center}
    \includegraphics[width=0.48\textwidth]{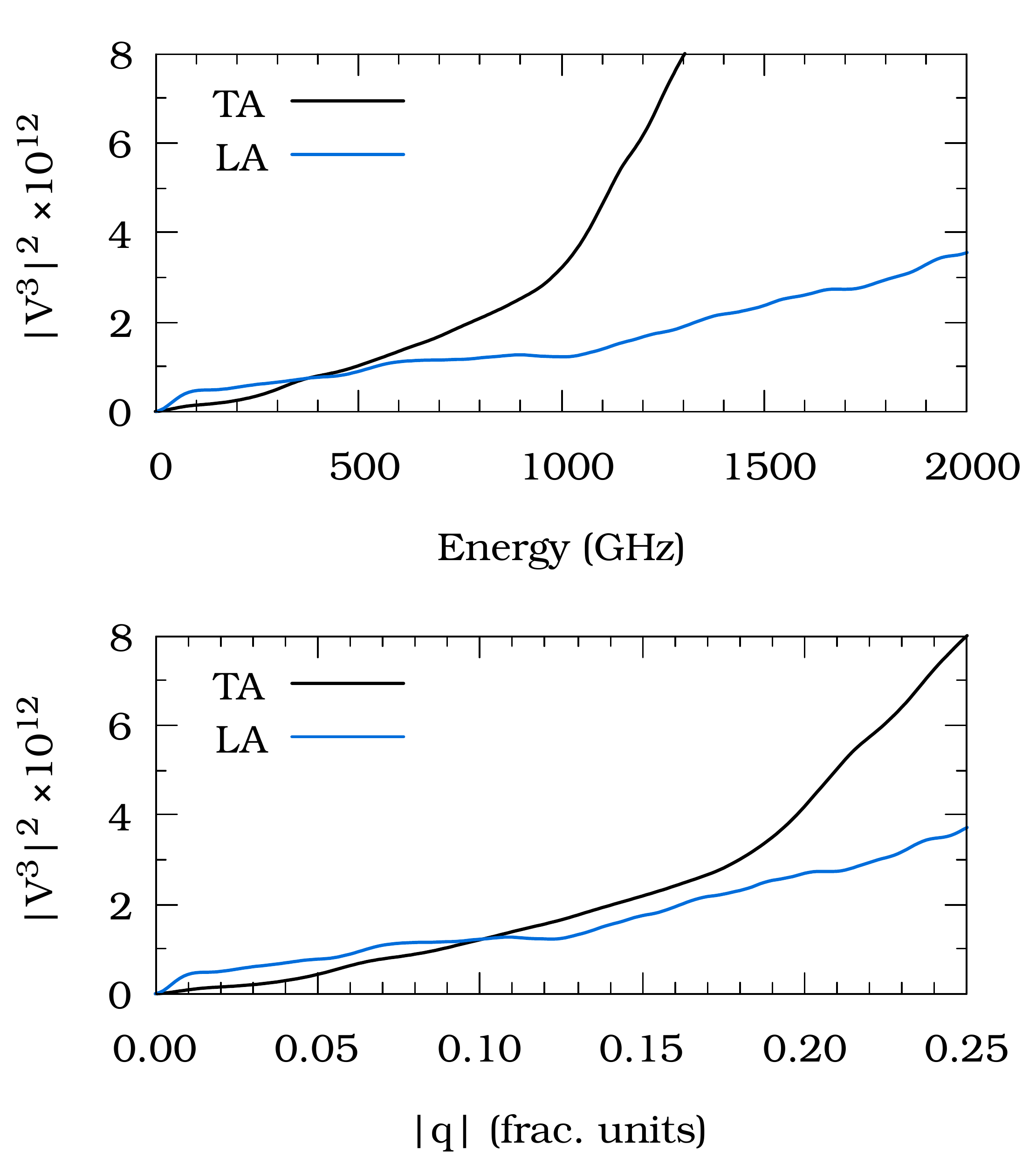}
  \end{center}
  \caption{GaAs, [100] direction. Squared matrix element involved in the scattering of the initial longitudinal (black  line) and transverse (blue line) phonons,  averaged over the BZ with respect to the $\mathbf{q'}$ wavevector, represented    as a function of $q_x$ component of the initial $\mathbf{q}$ wavevector (lower panel) and initial phonon frequency (upper panel). }
  \label{fig:melement}
\end{figure}

The importance of the effect of the matrix elements is illustrated in figure~\ref{fig:melement}, where we report the average value of $\left|V^{(3)}\right|^2$ involved in the scattering process of the  LA and TA phonons along the $[100]$ direction of GaAs, as a function of phonon frequency and wave vector.
One can see that the value is almost linear at low energy for TA phonons,  but deviates from linearity around 1000 GHz, becoming more steep. For LA phonons, on the contrary, the behavior of the average matrix element is not linear even below 1000 GHz, and exhibits, after an initial growth, 
a saturation between 600 and 1000 GHz, followed by a rapid increase above 1000 GHz, when processes other than Herring one start to dominate the scattering.
The convolution of $\left|V^(3)\right|^2$ with the T-JDOS explains most of the shape of the attenuation in function of $\omega$, with the sharp features being imputable to the complex interplay between the matrix-element magnitude and the overlap of phonon polarizations, which are inherent to the definition~\cite{Paulatto:2013} of $V^{(3)}$.

\section{Results for silicon}
\label{subsec:attenuation_si}

\begin{figure*}
  \subfloat[a][50~K]{\includegraphics[width=0.48\textwidth]{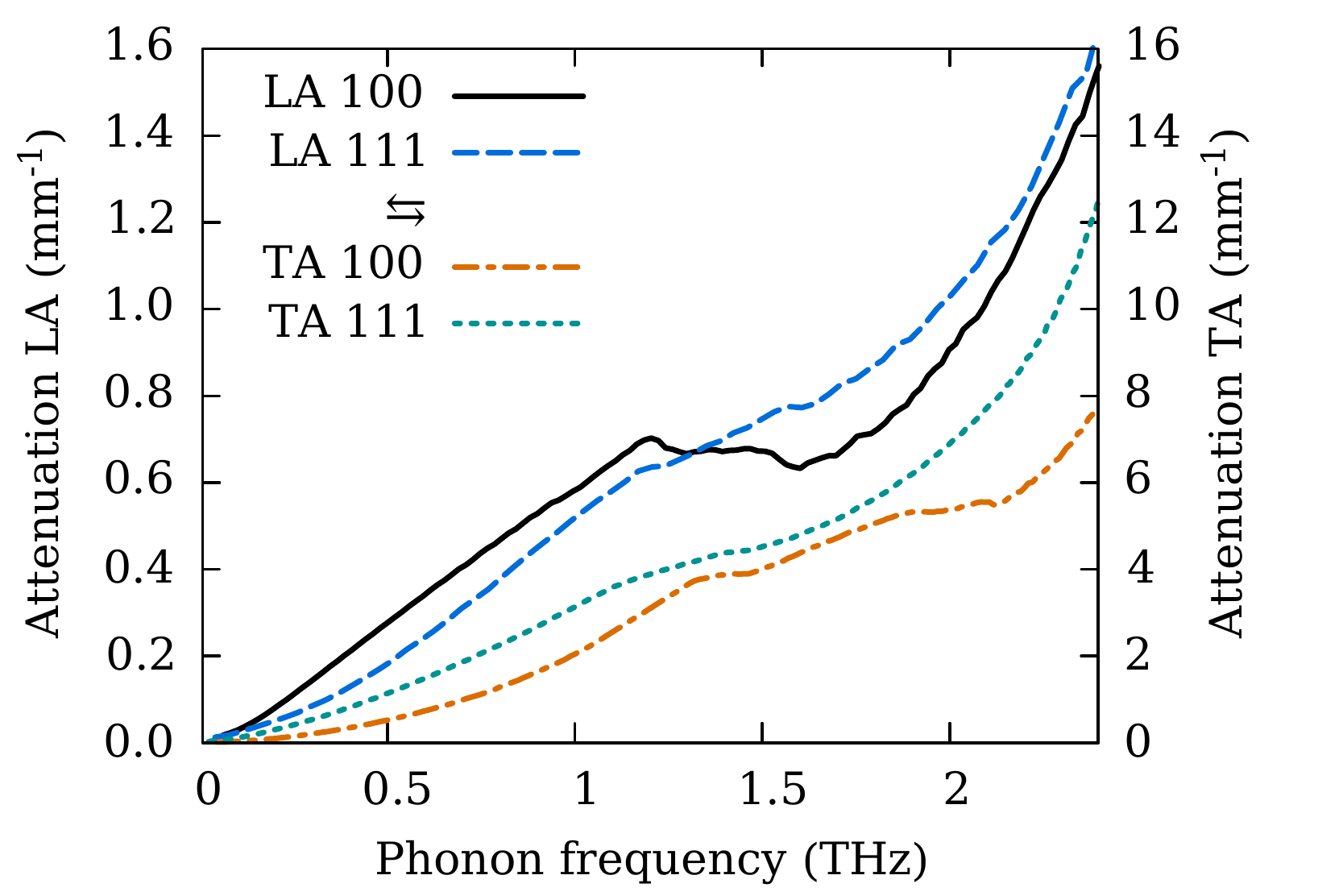}}
  \subfloat[b][300~K]{\includegraphics[width=0.48\textwidth]{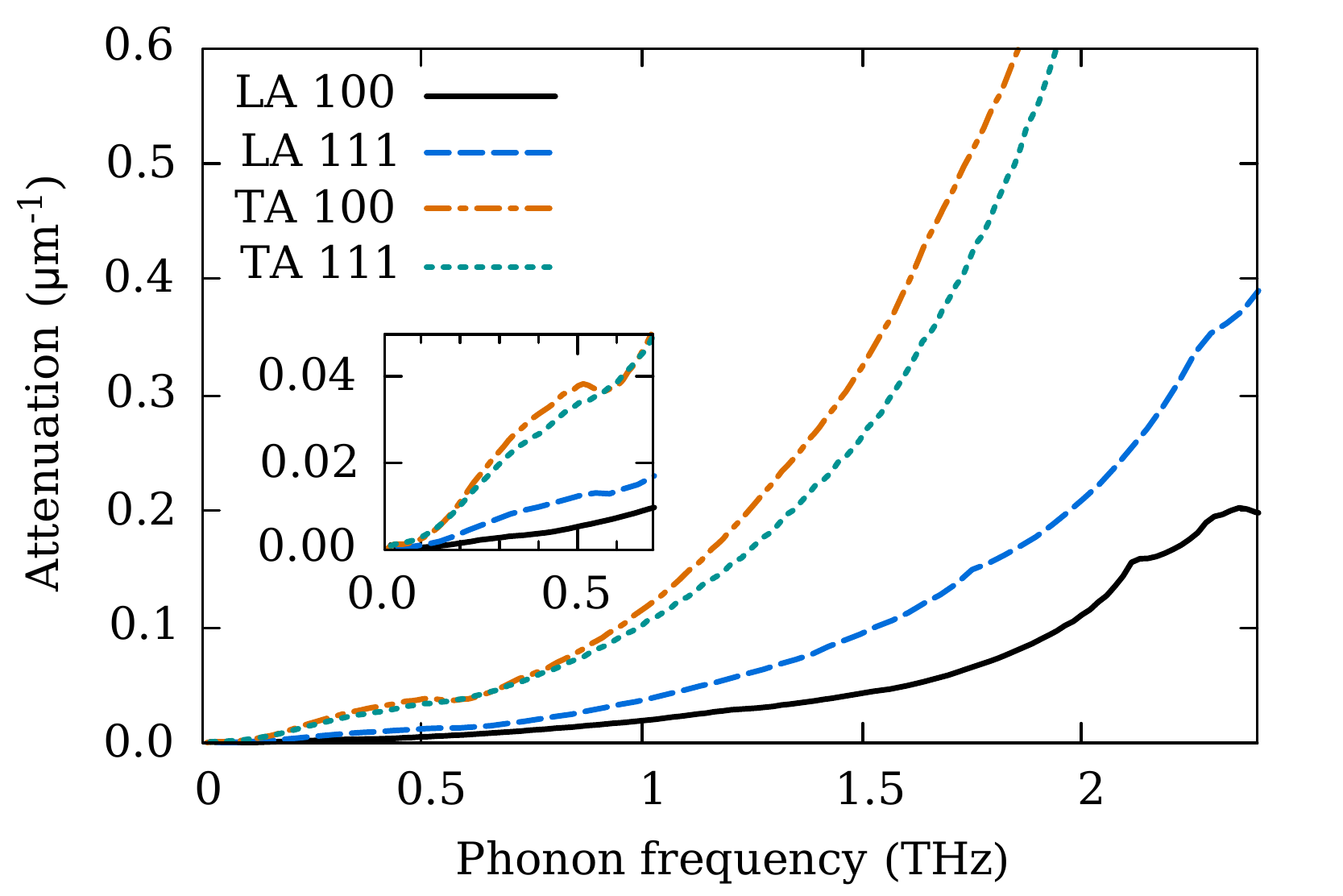}\label{fig:ofomega-si-b}}
  \caption{Silicon. Attenuation of the longitudinal and transverse acoustic phonons along the $[100]$ and $[111]$ directions at 50~K (left panel) and 300~K (right panel). Note the change of scale between the two temperatures (panels), as well as the different axis (on the right) for the TA mode in the left panel. }
  \label{fig:ofomega-si}
\end{figure*} 

In figure~\ref{fig:ofomega-si}, we report the computed attenuation of  LA and TA phonons along the $[100]$ and $[111]$ directions at 50~K and 300~K. 
At low temperature (panel a), there is a visible plateau in the LA and TA branches along the $[100]$ direction between 1.2~THz and 1.7~THz. When we examine the  $[111]$ direction, as in the case of GaAs, the plateau is not as pronounced as in the [100] direction. Nonetheless  a  change of behavior is clearly  visible in the same energy range as in the $[100]$ direction. 

At room temperature (panel b), the curves look smooth in the THz region. However, a careful examination reveals the presence of a small plateau, or a shoulder, around  500~GHz , which is magnified in the inset of figure \ref{fig:ofomega-si-b}.

We  conclude that our theoretical results predict a plateau for the attenuation of the LA [100] phonon between 1.2~THz and 1.7~THz, similar to the one which is found in GaAs both experimentally and theoretically.

To the best of our knowledge, there is no experimental data available for the attenuation
of acoustic waves in silicon in the range of frequencies for which we predict the attenuation plateau. There are few experimental results of sound absorption in silicon for frequencies above 10~GHz~\cite{Hao:2001,Daly:2009,Dekorsy:2009}. 
In Ref.~\onlinecite{Hao:2001}, data have been obtained in the 50-100~GHz frequency and 30-130~K temperature ranges.  As explained in the introduction, such conditions correspond to an intermediate regime coming from the transition between the LR regime and the Akhiezer one,  where the collective behavior of the phonon gas comes into play. The study of the intermediate regime is beyond the scope of the present work.

\begin{figure}
  \includegraphics[width=0.48\textwidth]{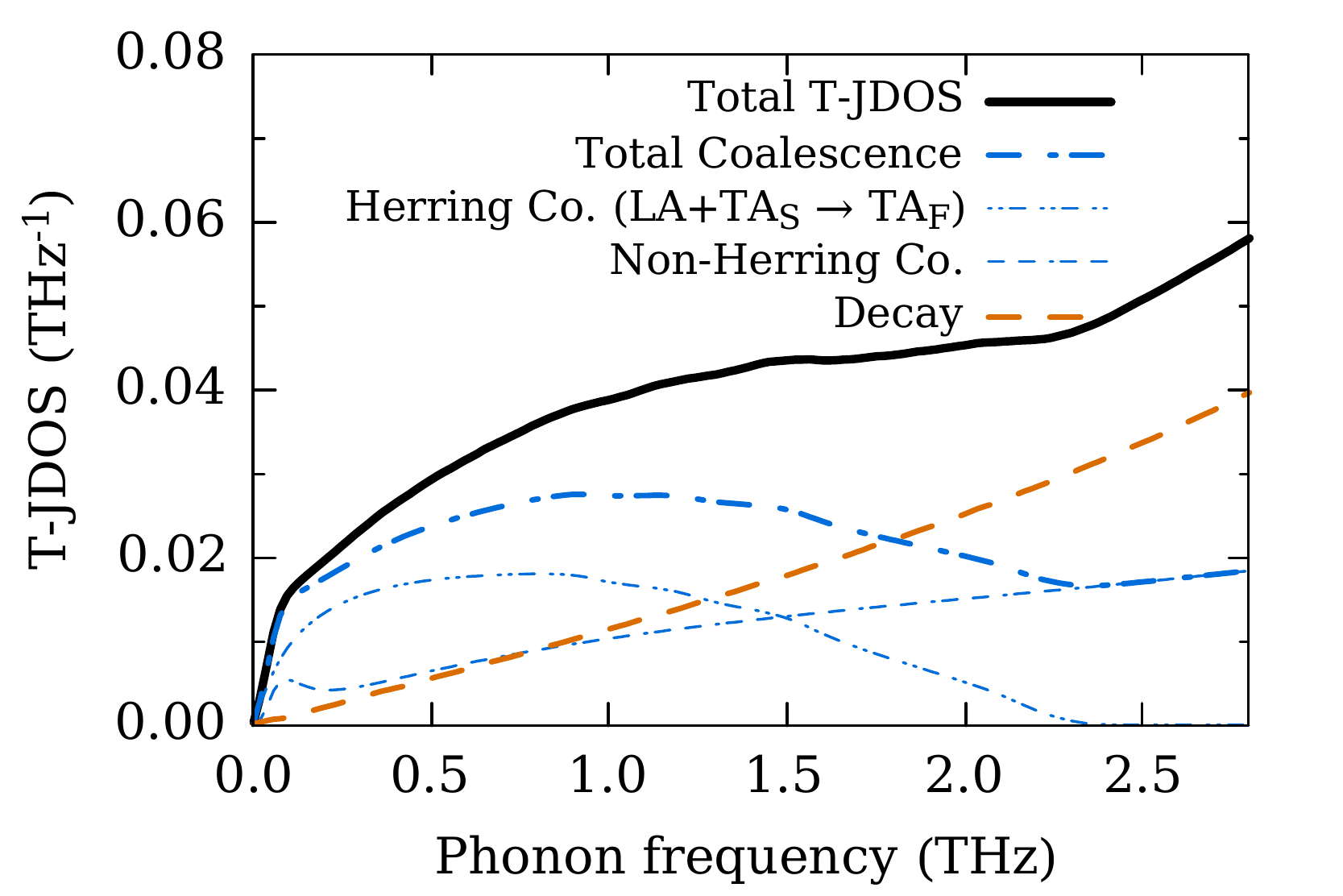}
\caption{Temperature-dependent joint-density of states (T-JDOS) at 50~K for the LA branch of silicon along the $[100]$ direction. The total T-JDOS (thick black solid line) is decomposed into  decay (dashed orange) and coalescence (dot-dashed blue) processes, the latter being further divided into Herring's (LA$\rightarrow$TA$_S+$TA$_F$, blue dash-dotted lines) and Non-Herring's (blue long-dashed lines) processes.}
\label{fig:ljdos-silicon}
\end{figure}

In analogy to what we have seen for GaAs in section~\ref{sec:origin-gaas}, we examine in figure~\ref{fig:ljdos-silicon} the T-JDOS of silicon. Herring's mechanism dominates in a frequency range similar to the one in GaAs, up to 1~THz, however the T-JDOS has a shorter and steeper onset and a larger saturation area, and it dies up slower than in GaAs, around 2.4~THz. This difference with respect to GaAs is consistent with the higher values of the phonon energy  and group velocity of silicon. As a consequence, at low temperatures, the plateau spans a larger energy range than in GaAs, and should be observable. 

At 300~K  (results not shown in the main text), the weight of scattering processes involving high energy phonons increases: a small plateau is visible  around 500\,MHz, and can be associated with the initial high steepness of the Herring T-JDOS. On the other hand, at 300~K, the large maximum of Herring's mechanism is not sufficiently steep to appear in the total attenuation curve, unless a careful examination is performed, as done in the inset of figure \ref{fig:ofomega-si-b}.  

\section{Generalization to other cubic semiconductors}
We have presented in detail how the breakdown of the Herring scattering model produces a plateau or a shoulder, and discussed the conditions of their observation in measurements of the attenuation $\alpha$ for two specific materials. In the following, we explain that it is a common feature of cubic semiconductors.

Indeed, Herring's model requires that the LA phonon along the [100] direction coalesces with a non-degenerate TA phonon into another TA phonon, the main regions that are active in the mechanisms are around the [100] and [111] directions (fig~\ref{fig:fstate-gaas}), with the former giving the main contribution at very low frequencies. The model assumes that all three phonons energies linearly depend  on their wavevector: $\omega_\nu = A_\nu |\qq_\nu|$ ($\nu=$LA, TA$_\mathrm{s}$, TA$_\mathrm{f}$). As a result, it is always possible to find a coalescence channel, as the phase space  is proportional to $|\qq|^2$. However, in order to have $\omega_\mathrm{LA} + \omega_\mathrm{TA_s} = \omega_\mathrm{TA_f}$, the two TA phonons have to be at an energy higher than LA, and they quickly exceed the region where their dispersion is  linear. The linear regime is in general quite limited in cubic semiconductors, up to $|q|\simeq 0.2 - 0.3$ inverse lattice units, in Si and GaAs.

A similar and complementary geometric argument holds for the main non Herring decay process, $\omega_\mathrm{LA} = \omega_\mathrm{TA_\mathrm{fast}} + \omega_\mathrm{TA_\mathrm{slow}}$:  the  phase-space for decay is initially very small, being limited by energy conservation, and increases with $\omega_\mathrm{LA}$.

The total attenuation is initially the sum of two curves, and its behavior is an interplay between the two contributions: the Herring coalescence one that increases, saturates and then decreases; and the non-Herring decay one, that increases from zero. It is clear that because of the different behaviors,  a plateau will form, although its exact energy range and amplitude depend on the specific material, and can be predicted, as we show in the present work, by ab initio calculations.

\section{Model based on elastic constants}

\begin{figure}
  \begin{center}
    \includegraphics[width=0.48\textwidth]{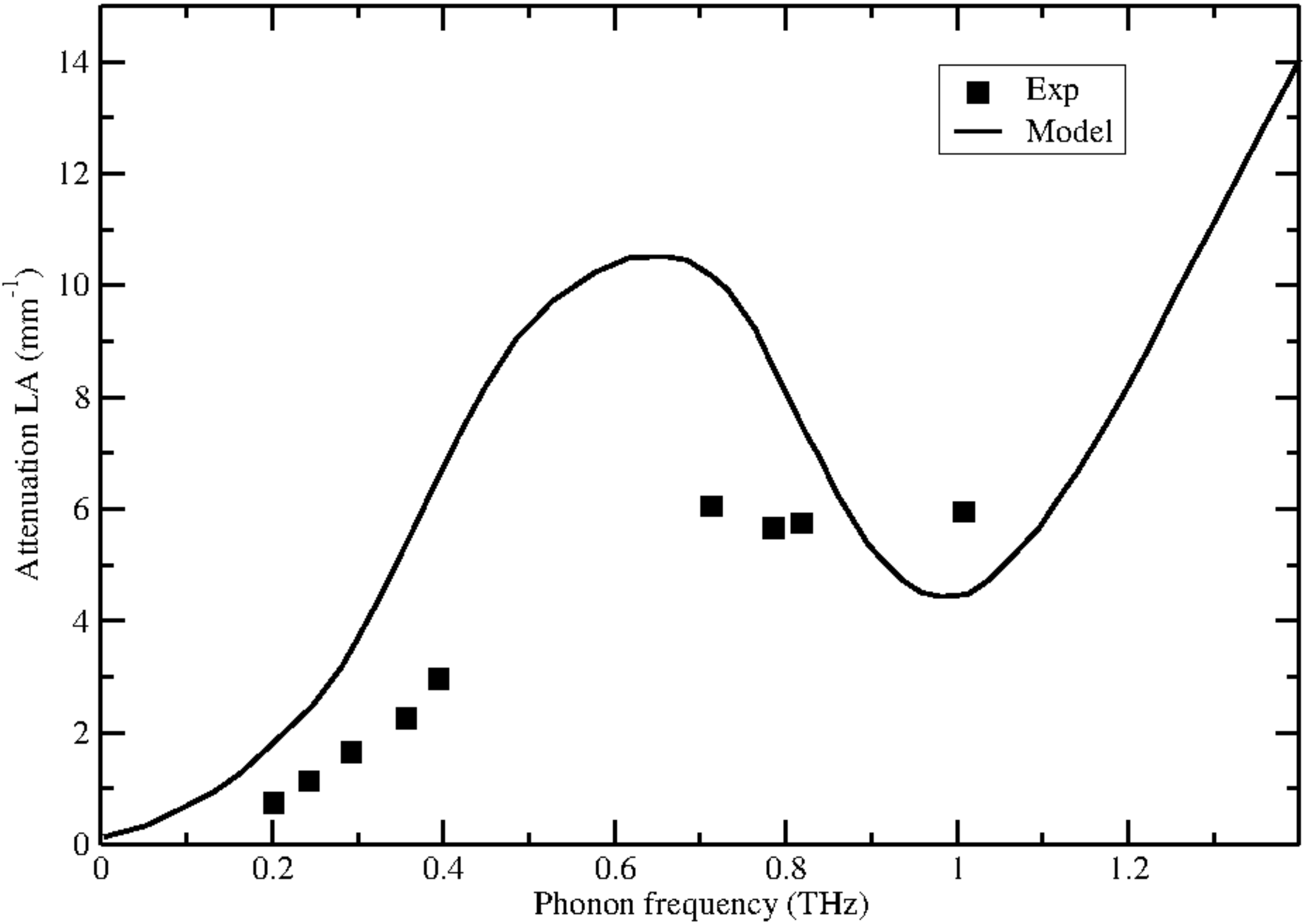}
  \end{center}
  \caption{GaAs at 50 K. Lines: attenuation of LA phonon obtained in  Ref.~\onlinecite{Legrand:2014} with a 
  modified model based on elastic constants and \emph{ab initio} phonon dispersion (see text). 
  Symbols: experimental data of Ref.~\onlinecite{Legrand:2016}.
  }
  \label{fig:romain}
\end{figure}

It is worth noting that the information about the behaviour of the phonon-phonon matrix elements discussed above, can be used to evaluate the applicability range of models used in the past \cite{Simons:1957,Tamura:1985,Berke:1988} to describe
the phonon-phonon interaction in materials. In these models,  phonon-phonon matrix elements are described \emph{via} linear combinations of elastic constants,
and are linearly proportional to the phonon wavevectors of the three phonons involved in the interaction process (see Appendix for details).   
As one can see from Fig.~\ref{fig:melement}, the behavior of the average of the matrix elements for  phonon-phonon interaction starts to deviate more and more from  linearity 
for both transverse and longitudinal phonons as the modulus of the wavector grows. We have shown above in Fig.~\ref{fig:fstate-gaas} that the phonons involved in three-phonon 
scattering processes responsible for acoustic wave attenuation can have wavevectors close to the BZ boundary. Indeed, it was shown in an earlier work~\cite{Legrand:2014} that it is impossible
to reproduce the experimentally observed plateau in the attenuation of the LA $[100]$ phonon in GaAs using the unmodified model based on 
 elastic constants for the phonon-phonon interaction matrix elements. However, it was also shown in Ref.~\onlinecite{Legrand:2014} that if
the model for the phonon-phonon interaction matrix element was modified to account for saturation of the matrix elements (with an adjustable parameter), and the \emph{ab initio} phonon dispersion
was used, then the plateau could be  reproduced, as  shown in Fig.~\ref{fig:romain}. Indeed, the qualitative similarity of the theoretical results obtained in Ref.~\onlinecite{Legrand:2014} 
with a (modified) model based on elastic constants  and \emph{ab initio} phonon dispersion, shown in Fig.~\ref{fig:romain},  to the ones presented in this work 
in Fig.~\ref{fig:ofomega}, highlights once
again the fact that the plateau is due to the interplay between T-JDOS (which is determined solely by the phonon dispersion) and 
the phonon-phonon matrix elements. One must note, however, that quantitatively, the fully \emph{ab initio} results 
of Fig.~\ref{fig:ofomega} agree much better with experimental data than the ones presented on Fig.~\ref{fig:romain}. 
The details of the model proposed in work~\cite{Legrand:2014}  are described in the Appendix.

\section{Conclusions}
We have computed within density functional perturbation theory the intrinsic phonon-phonon scattering processes for the very low frequency region of crystalline gallium arsenide and  silicon phonon dispersion. For GaAs we have compared the calculations along the $[100]$ direction  with experiments, finding the agreement to be excellent, both as a function of the phonon frequency, at constant temperature, and as a function of temperature, for a given phonon frequency.

We are able to give some insight into the mechanisms underlying the anomalous, but quite general, appearance of a plateau (especially for the LA phonon) or shoulder (for TA) in the phonon attenuation as a function of frequency, appearing between 600~GHz and 1~THz for GaAs, and between 1.2 and 1.7~THz for silicon. 

Specifically, the plateau is caused by the 3-phonon coalescence Herring processes, which dominate at low frequency, progressively saturating and then rapidly decreasing as the states involved in the scattering move up in energy, over the limit of the acoustic part of the phonon dispersion. The saturation is caused by the acoustic phonon dispersion changing from linear to constant at low energy, which is a common feature of cubic semiconductors.

One of the major consequences of the breakdown of Herring’s processes is that the absorption length of longitudinal acoustic waves at 1 THz in semiconductors could be pretty large even at room temperature (50$\mu$m at 300K in silicon for a wavelength of 8.5 nm), opening interesting possibilities for high resolution phonon imaging of deeply embedded nanostructures.

It is interesting to observe that the magnitude of the attenuation can be quite different for different directions and branches, but the plateau is still present grossly in the same frequency range. On the other hand, it becomes less visible at higher temperatures as more scattering channels involving the optical bands reduce the dominance of the simple Herring decay mechanism, drowning it out at lower frequencies.

\section{Acknowldegments}
This work was granted HPC resources of IDRIS, CINES and TGCC under the GENCI projects 7320 and 2210, and by the Ecole Polytechnique through the LLR-LSI project. We acknowledge support from the Chaire \'Energie of the \'Ecole Polytechnique.

\section*{Appendix}

We briefly present the model for the anharmonic coefficients and its modification proposed in Ref.~\onlinecite{Legrand:2014}.

In the long range approximation, the phonon-phonon matrix element can be written as \cite{Berke:1988}:
\begin{equation}
V^{(3)}(\qq j,\qq' j', \qq'' j'')=\frac{\Phi(\qq j,\qq' j', \qq'' j'')}{\omega_{j}\omega_{j'}\omega{j''}} \, ,
\end{equation}
with 
\begin{widetext}
\begin{align}
\begin{split}
&\Phi(\mathbf{q}_{j},\mathbf{q}_{j'},\textbf{q}_{j''}) = \displaystyle\sum_{\alpha\beta\gamma\delta\mu\nu}\mathbf{S}_{\alpha\beta,\gamma\delta,\mu\nu}\times 
\mathbf{q}_{j,\alpha} \mathbf{q}_{j',\gamma} \mathbf{q}_{j'',\mu}\times\mathbf{e}_{j,\beta}\mathbf{e}_{j',\delta}\textbf{e}_{j'',\nu}\, ,\\
&\mathbf{S}_{\alpha\beta,\mu\nu,\zeta\xi}=\textbf{C}_{\alpha\beta,\mu\nu,\zeta\xi}+\delta_{\alpha\mu}\mathbf{C}_{\beta\nu,\zeta\xi}+\delta_{\alpha\zeta}\mathbf{C}_{\mu\nu,\beta\xi}+\delta_{\mu\zeta}\mathbf{C}_{\alpha\beta,\nu\xi} \, ,
\end{split}
\label{long-range-model}
\end{align}
\end{widetext}
where $\mathbf{C}$ stands for both the second- and third-order tensors of elastic constants, $\mathbf{e}$ are the phonon polarisations.
As one can see, the linear dependence on the $\mathbf{q}$ vectors of the three phonons involved in the
interaction is built in the model of Ref.~\onlinecite{Berke:1988}. 
Thus, the model cannot describe the saturation of phonon frequencies and phonon-phonon matrix elements for phonon wavevectors
far from the center of the Brillouin zone. 

In order to take the saturation into account, a simple modification was proposed in Ref.~\onlinecite{Legrand:2014}:
\begin{widetext}
\begin{align}
\begin{split}
&\Phi(\textbf{q}_{j},\textbf{q}_{j'},\textbf{q}_{j''}) = \displaystyle\sum_{\alpha\beta\gamma\delta\mu\nu}\textbf{S}_{\alpha\beta,\gamma\delta,\mu\nu}\times 
\tilde{\mathbf{q}}_{j,\alpha} \tilde{\mathbf{q}}_{j',\gamma} \tilde{\mathbf{q}}_{j'',\mu}\times\textbf{e}_{j,\beta}\textbf{e}_{j',\delta}\textbf{e}_{j'',\nu}\\
&\tilde{\textbf{q}} = \begin{cases}
\textbf{q}\text{~~~~~~~~~~if~~} \vert \textbf{q} \vert <q_{c}\\
~q_c \cdot \dfrac{\textbf{q}}{\vert\textbf{q}\vert}\text{~~if~~} \vert \textbf{q} \vert\geq{q}_{c}\\
\end{cases}\\
&\textbf{S}_{\alpha\beta,\mu\nu,\zeta\xi}=\textbf{C}_{\alpha\beta,\mu\nu,\zeta\xi}+\delta_{\alpha\mu}\textbf{C}_{\beta\nu,\zeta\xi}+\delta_{\alpha\zeta}\textbf{C}_{\mu\nu,\beta\xi}+\delta_{\mu\zeta}\textbf{C}_{\alpha\beta,\nu\xi}
\end{split}
\end{align}
\end{widetext}

As one can see, the matrix elements are now allowed to grow linearly with $\mathbf{q}$ 
only until the $\mathbf{q}$-vector reaches a "saturation wavector" $q_c$. In Ref.~\onlinecite{Legrand:2014},
$q_c$ was treated as a fitting parameter, which we found to be $q_c=0.37 (\frac{2\pi}{a_0})$ for GaAs, for all the temperatures and all initial acoustic phonon frequencies
considered.

\bibliography{bibliography}

\end{document}


\title{Supplementary materials\\Breakdown of Herring's processes in cubic semiconductors for sub-terahertz longitudinal acoustic phonons}

\author{Maxime Markov}
\author{Jelena Sjakste}
\author{Nathalie Vast}
\affiliation{Laboratoire des Solides Irradi\'{e}s (LSI), \'Ecole Polytechnique - CEA/DRF/IRAMIS - CNRS UMR 7642,  91128 Palaiseau c\'edex, France}
\author{Romain Legrand}
\author{Bernard Perrin}
\affiliation{Institut de Nanosciences de Paris (INSP), Sorbonne Université, CNRS UMR 
7588, Case 840, 4 place Jussieu, 75252 Paris Cedex 05, France}
\author{Lorenzo Paulatto}
\email{lorenzo.paulatto@sorbonne-universite.fr}
\affiliation{Institut de Min\'eralogie, de Physique des Mat\'eriaux et de 
Cosmochimie (IMPMC), Sorbonne Université, CNRS UMR 
7590, IRD UMR 206, Case 115, 4 place Jussieu, 75252 Paris Cedex 05, France}

\maketitle{}


\begin{figure}[htb]
\subfloat[a][202\,GHz]{\includegraphics[width=0.48\textwidth]{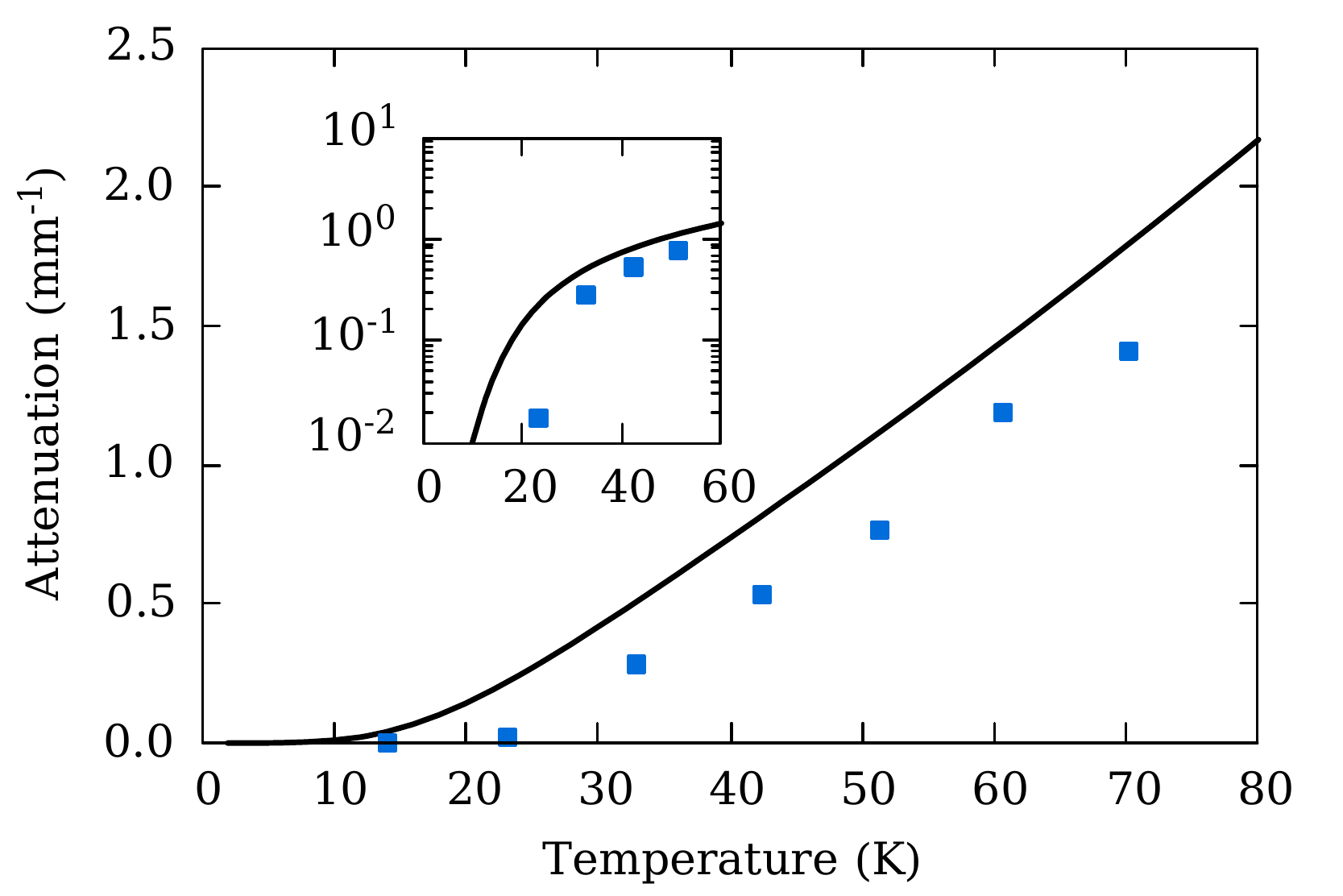}}
\subfloat[b][243\,GHz]{\includegraphics[width=0.48\textwidth]{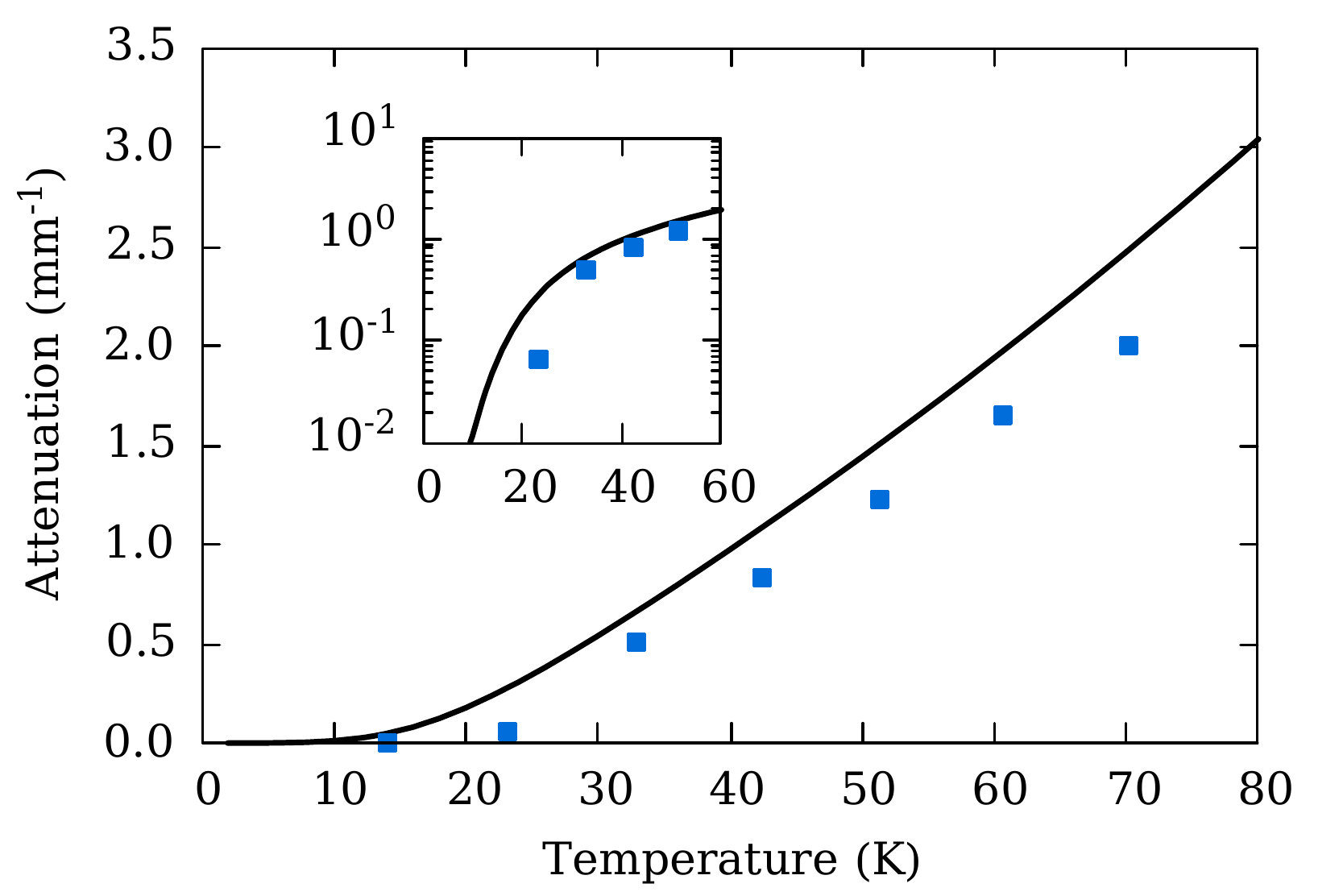}}
\\
\subfloat[c][292\,GHz]{\includegraphics[width=0.48\textwidth]{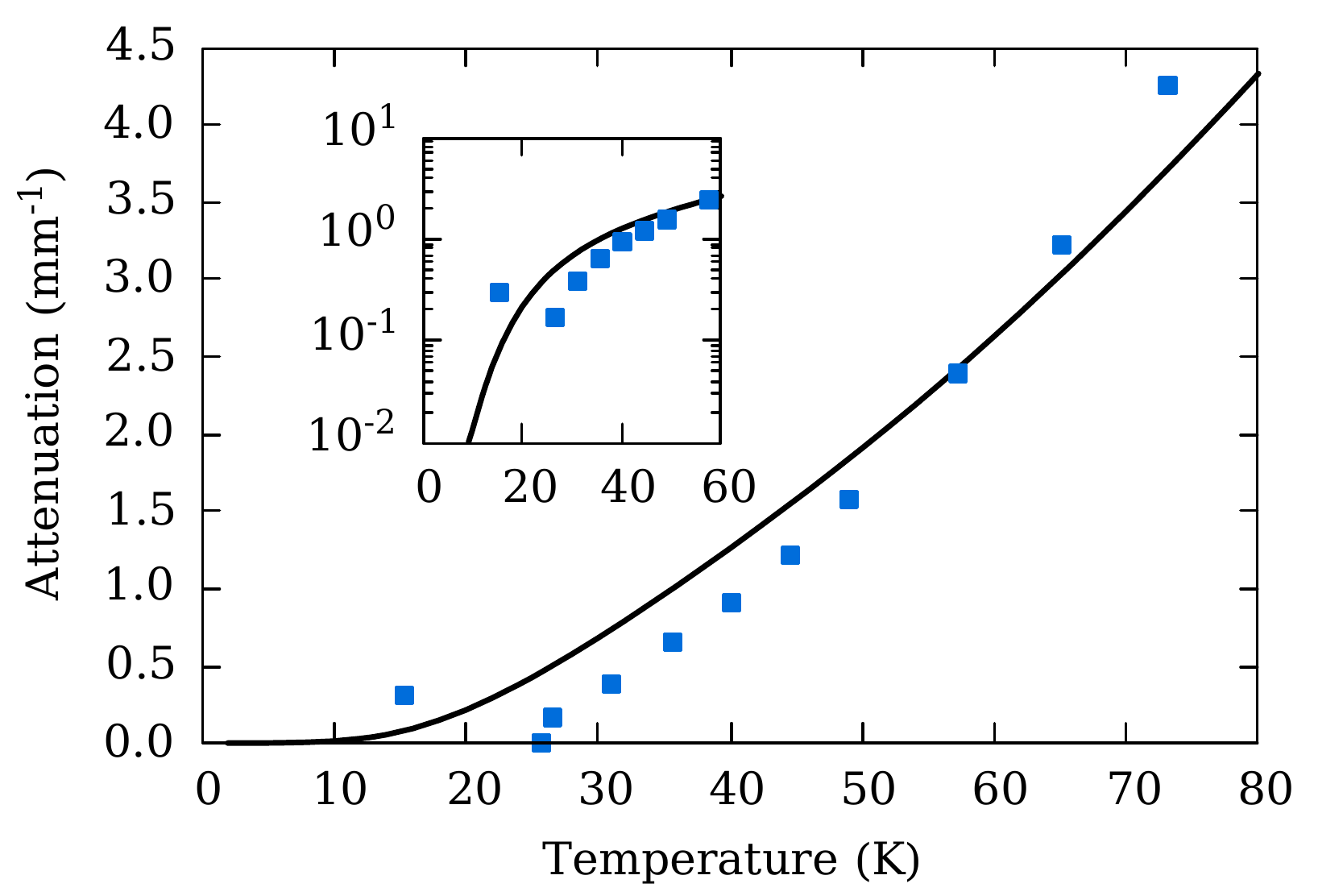}}
\subfloat[d][356\,GHz]{\includegraphics[width=0.48\textwidth]{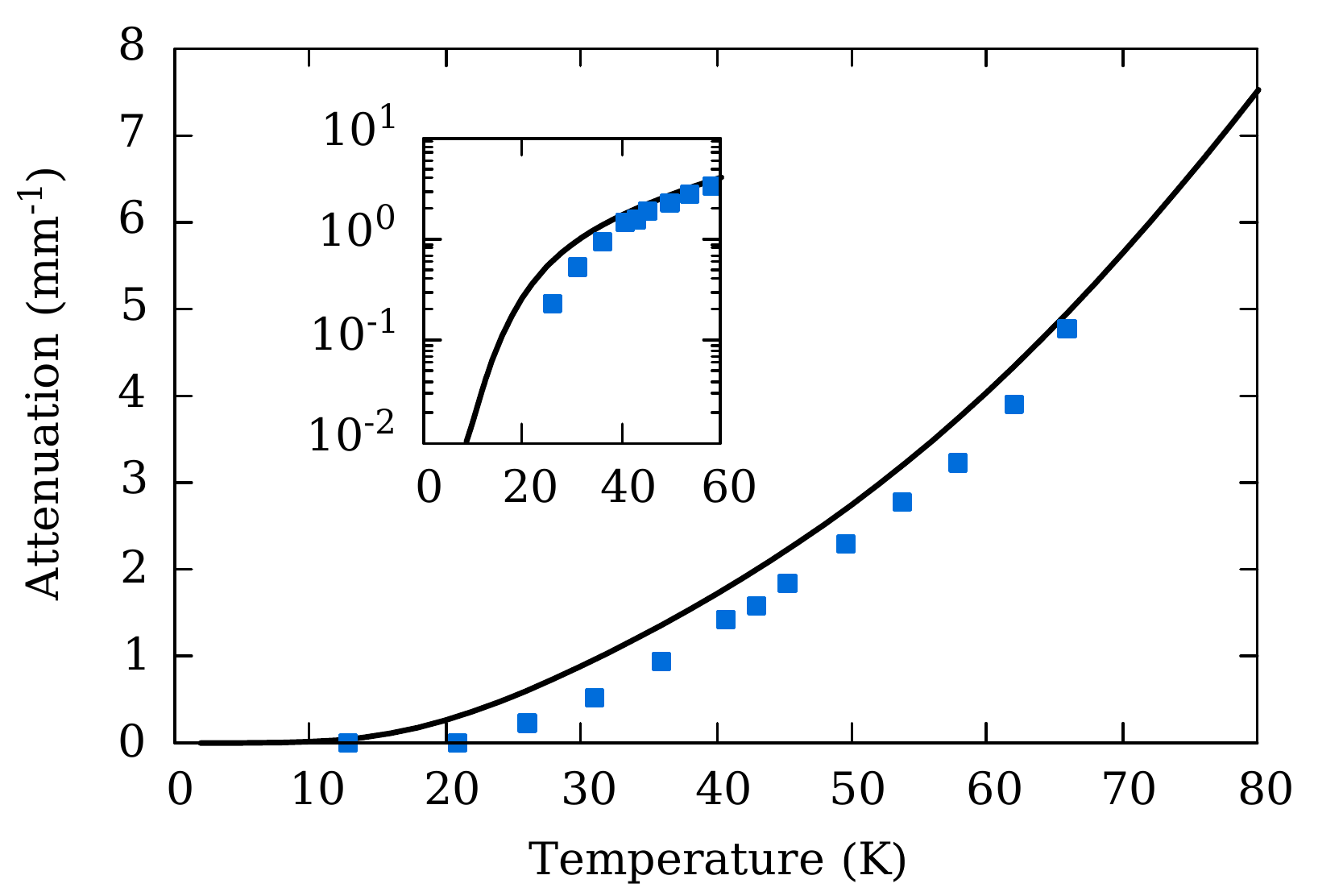}}
\caption{Attenuation of LA phonons in GaAs along {$[100]$} at different frequencies of the initial phonon, as a function of the temperature. Theoretical results (black line) are compared with experiments\cite{Legrand:2016}. Note that experimental data take the smallest value (typically the one at the lowest temperature) as reference, in order to cancel out intrinsic scattering sources (isotopes, defects, boundaries), hence they will go to zero at a finite temperature, contrary to the theoretical curve that has a finite value at 0~K (see main text).}
\label{fig:oft-supp}
\end{figure}
\begin{figure}[htb]
\subfloat[e][394\,GHz]{\includegraphics[width=0.48\textwidth]{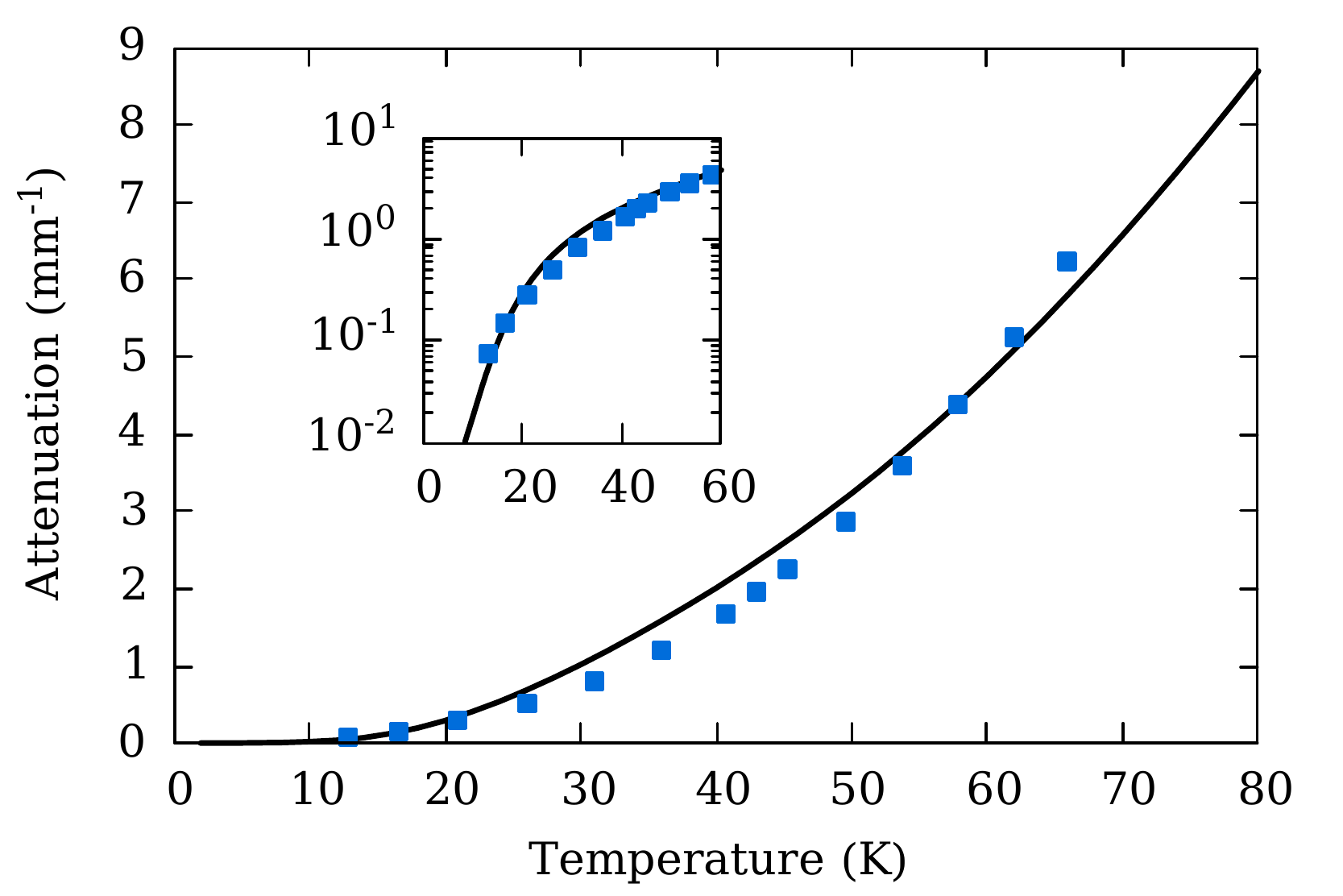}}
\subfloat[f][786\,GHz]{\includegraphics[width=0.48\textwidth]{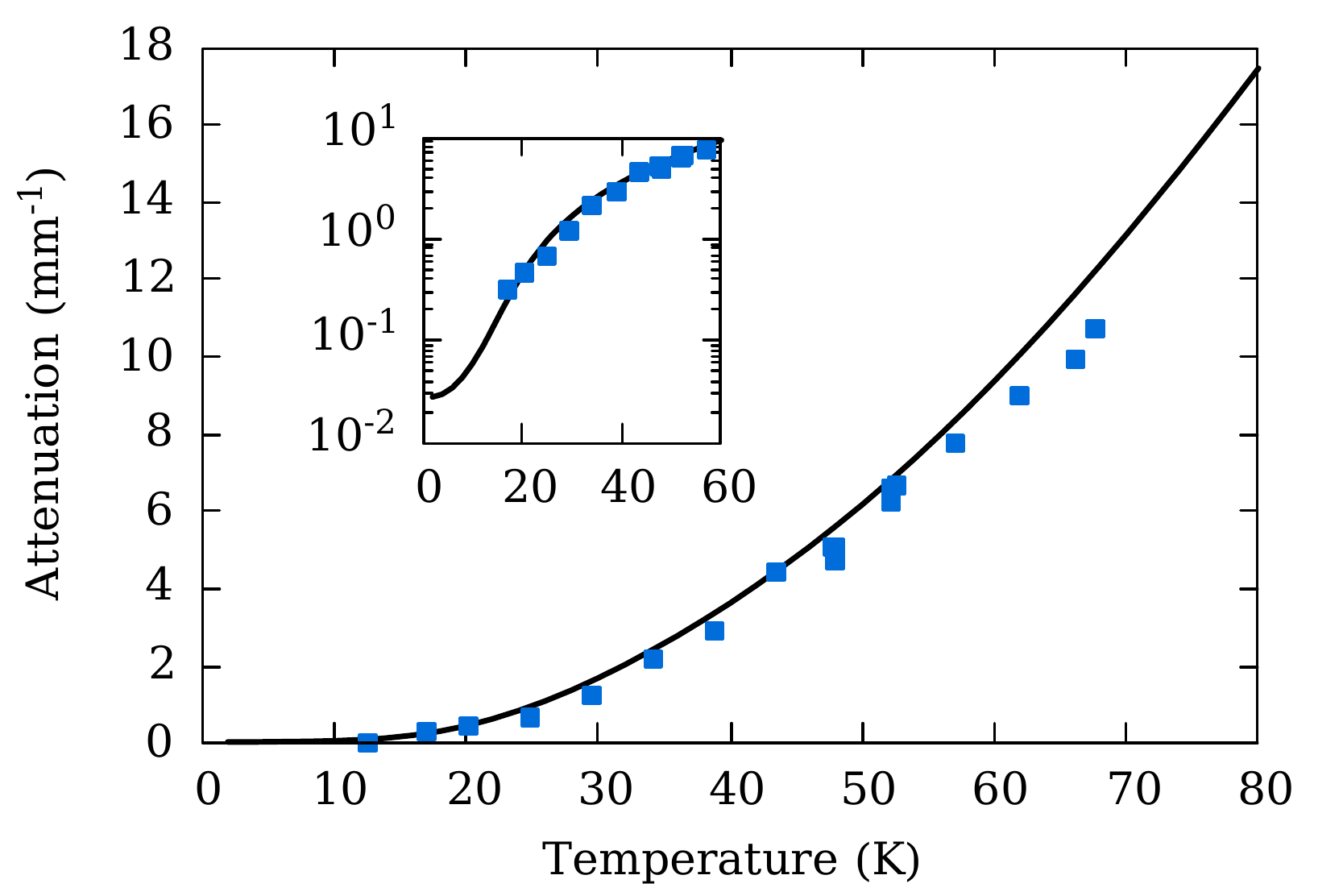}}
\\
\subfloat[g][819\,GHz]{\includegraphics[width=0.48\textwidth]{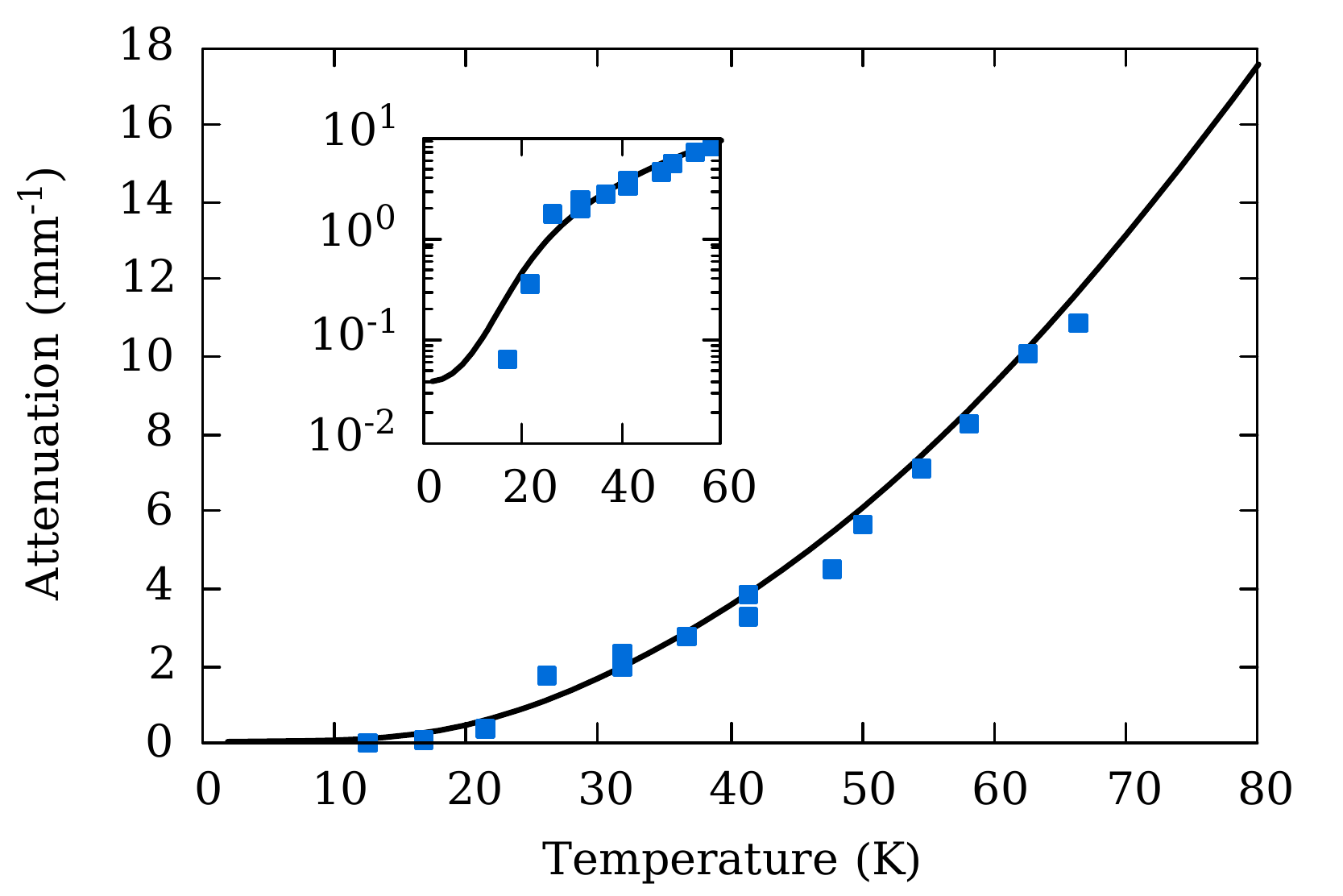}}
\subfloat[h][1000\,GHz]{\includegraphics[width=0.48\textwidth]{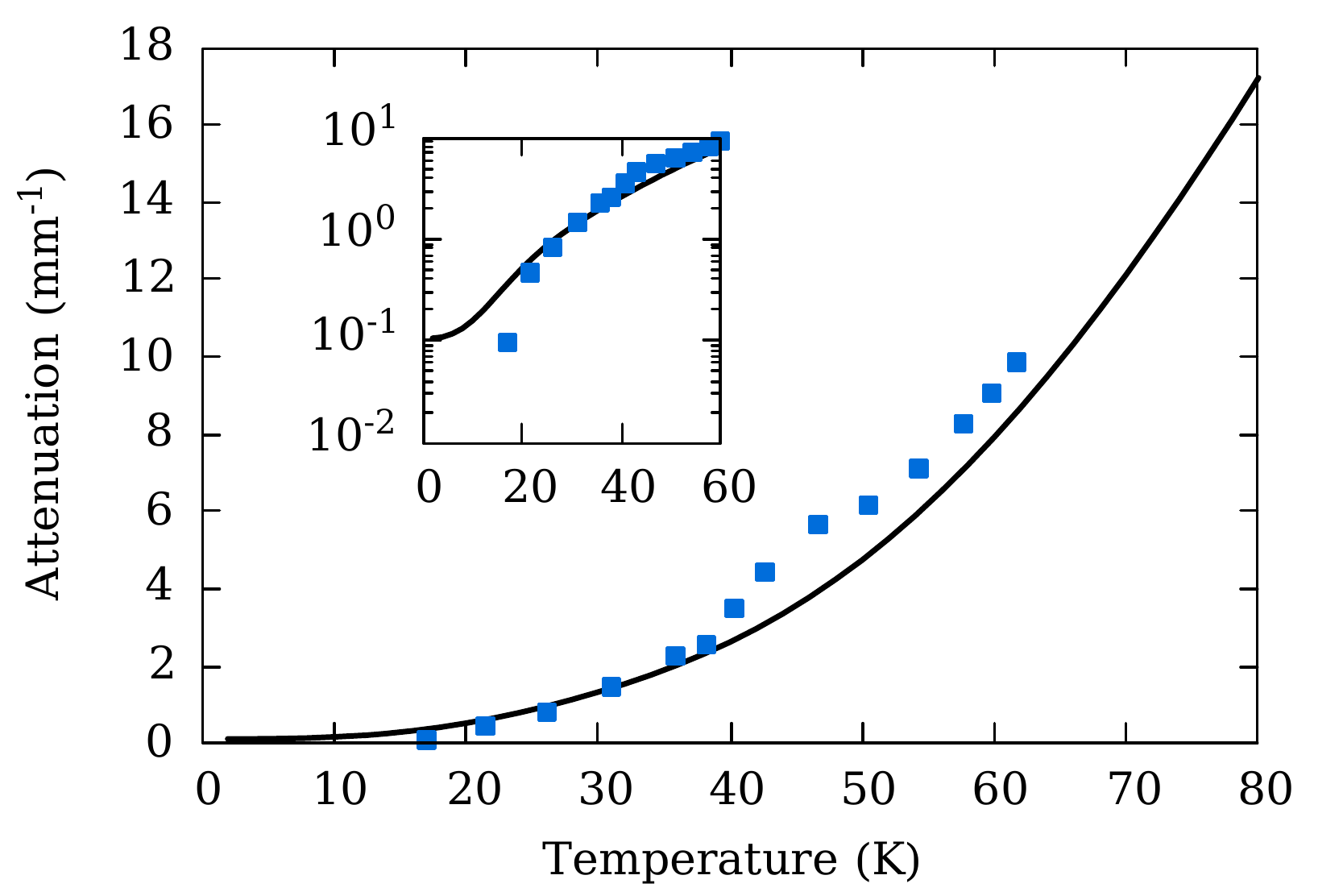}}
\caption{Attenuation of LA phonons in GaAs along the [100] direction at 394, 786, 819 and 1000 GHz. Same comment as figure 1.}
\end{figure}


\begin{figure}[htb]
\subfloat[a][GaAs {$[100]$} 50\,K]{\includegraphics[width=0.48\textwidth]{gaas-jdos-50K.pdf}}
\subfloat[b][GaAs {$[100]$} 300\,K]{\includegraphics[width=0.48\textwidth]{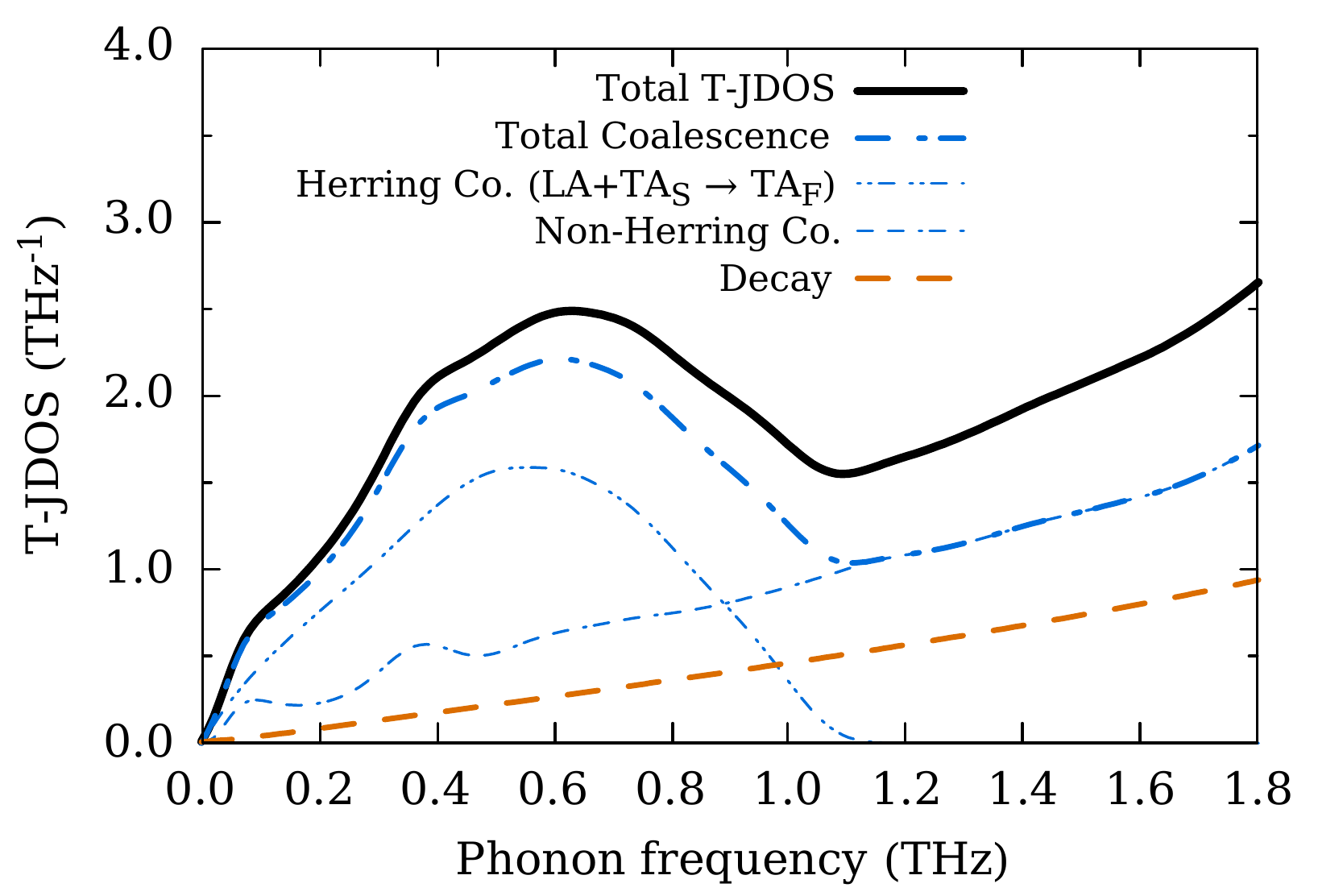}}
\\
\subfloat[c][GaAs {$[111]$} 50\,K]{\includegraphics[width=0.48\textwidth]{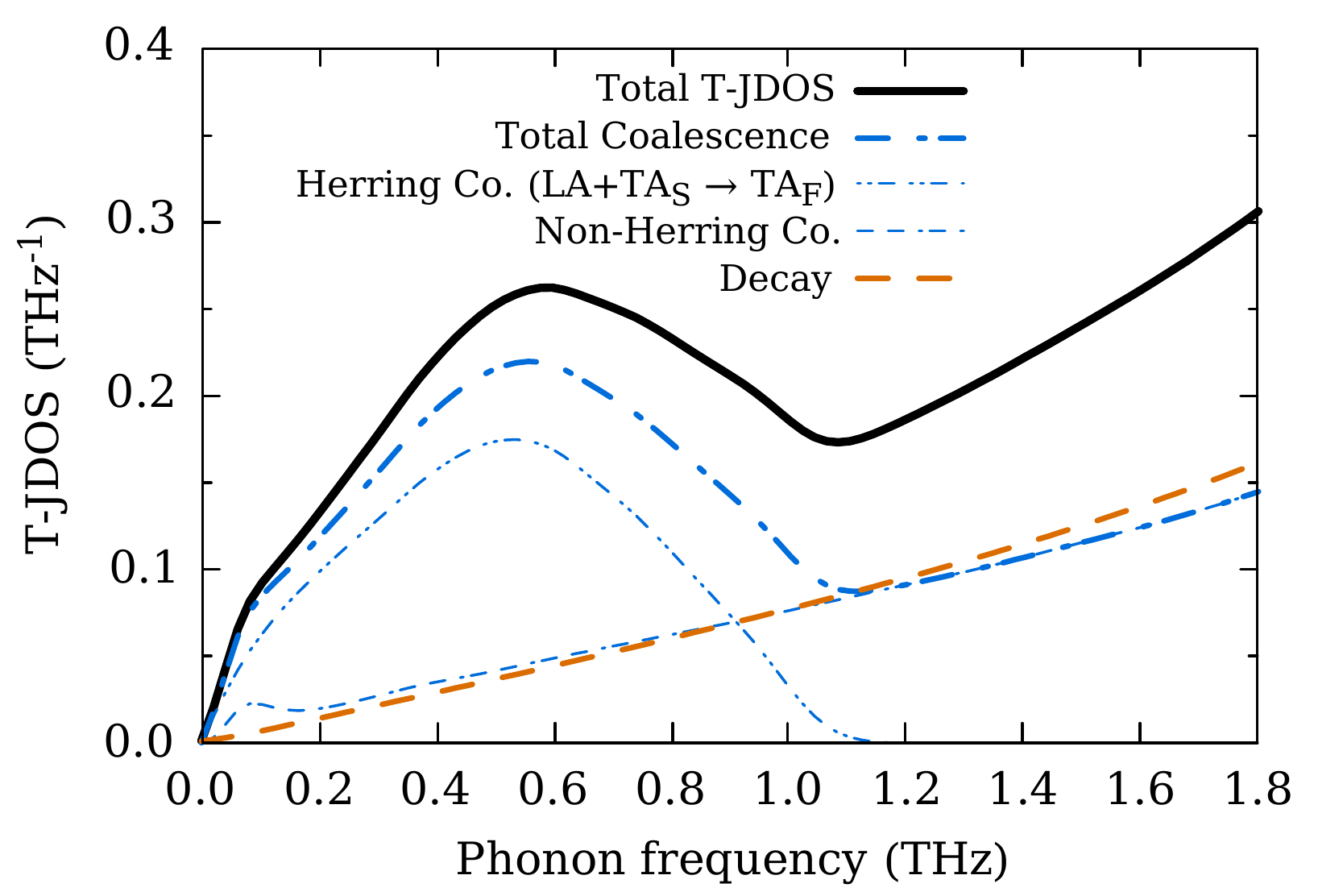}}
\subfloat[d][GaAs {$[111]$} 300\,K]{\includegraphics[width=0.48\textwidth]{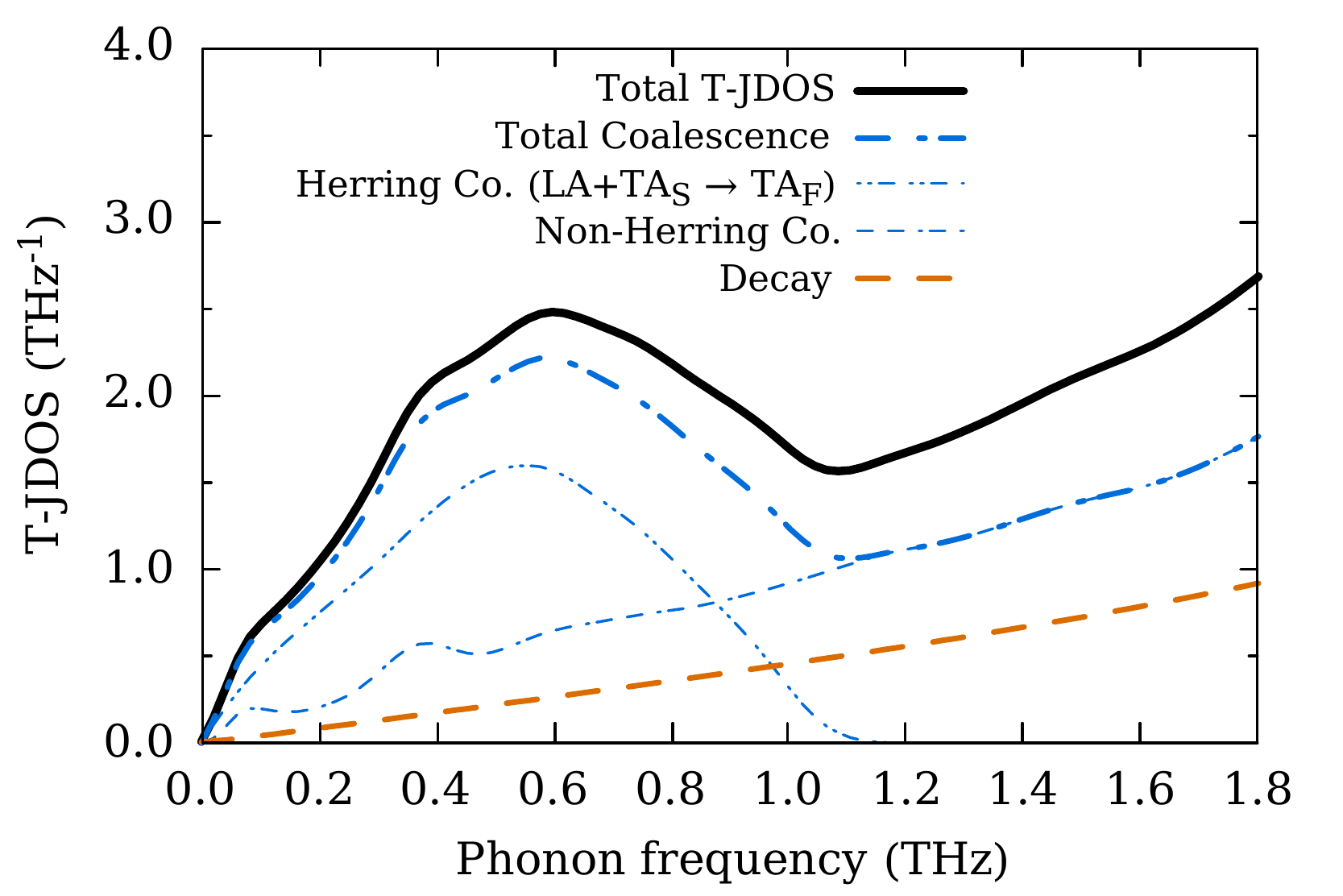}}
\caption{temperature-dependent joint phonon density of states (T-JDOS) for GaAs at 50\,K along directions {$[100]$} and {$[111]$}. We underline: 1) The very high isotropy of these L-JDOS. 2) The effect of temperature which only changes the position of the maximum of Herring processes and the relative intensity of scattering and coalescence processes.}
  \label{fig:ljdos-supp}
\end{figure}
\begin{figure}[htb]
\subfloat[e][Silicon {$[100]$} 50\,K]{\includegraphics[width=0.48\textwidth]{si-jdos-50K.pdf}}
\subfloat[f][Silicon {$[100]$} 300\,K]{\includegraphics[width=0.48\textwidth]{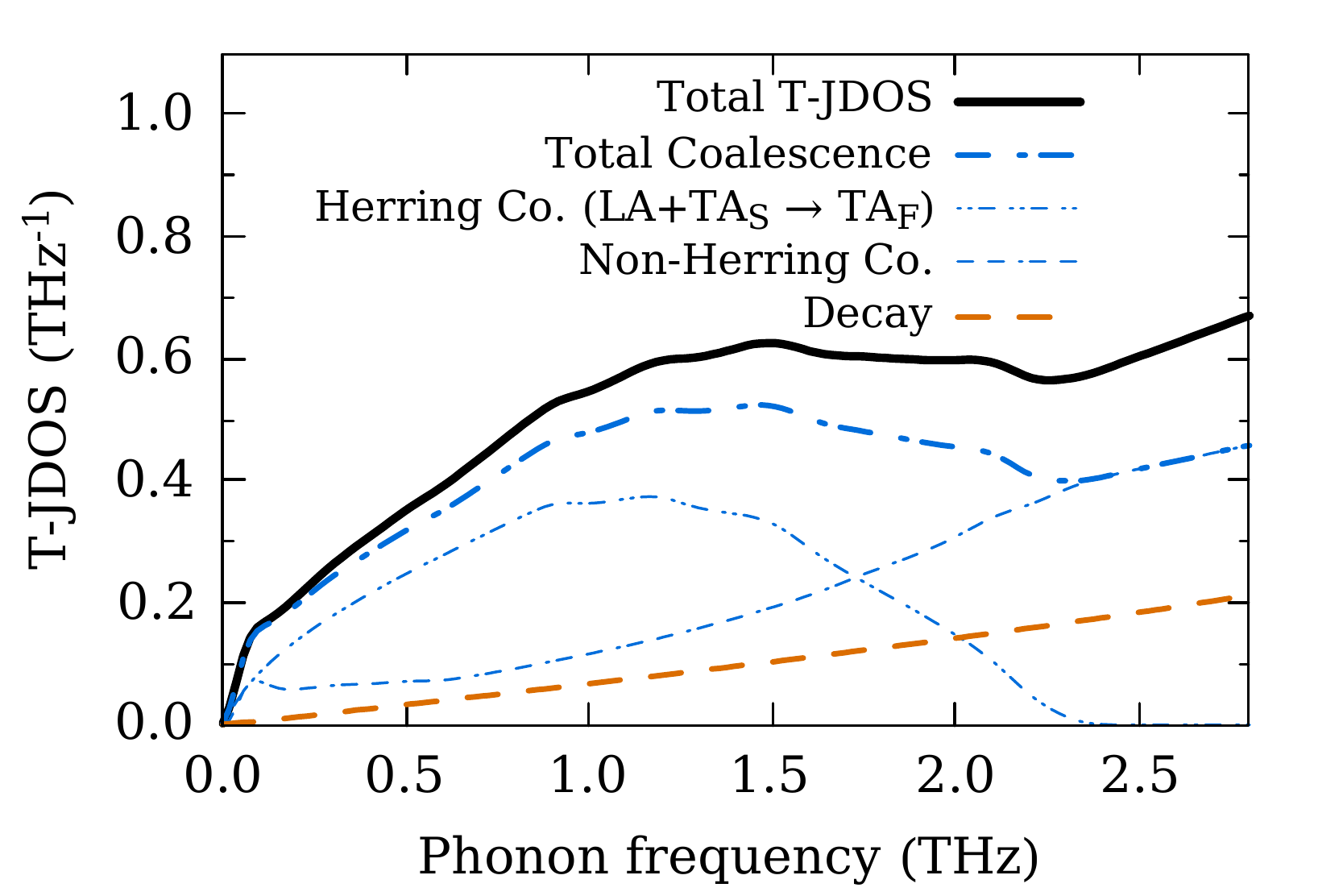}}
\\
\subfloat[g][Silicon {$[111]$} 50\,K]{\includegraphics[width=0.48\textwidth]{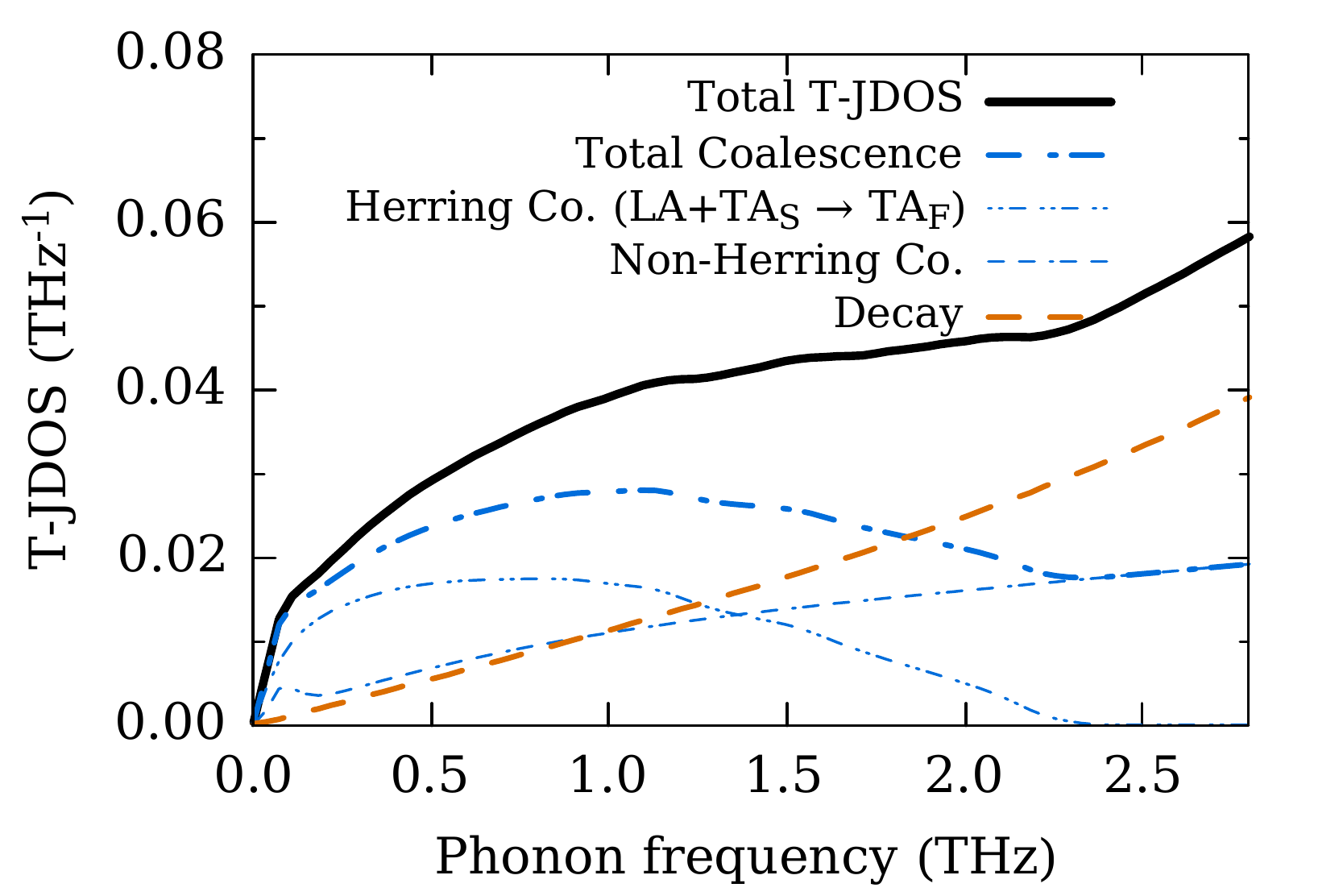}}
\subfloat[h][Silicon {$[111]$} 300\,K]{\includegraphics[width=0.48\textwidth]{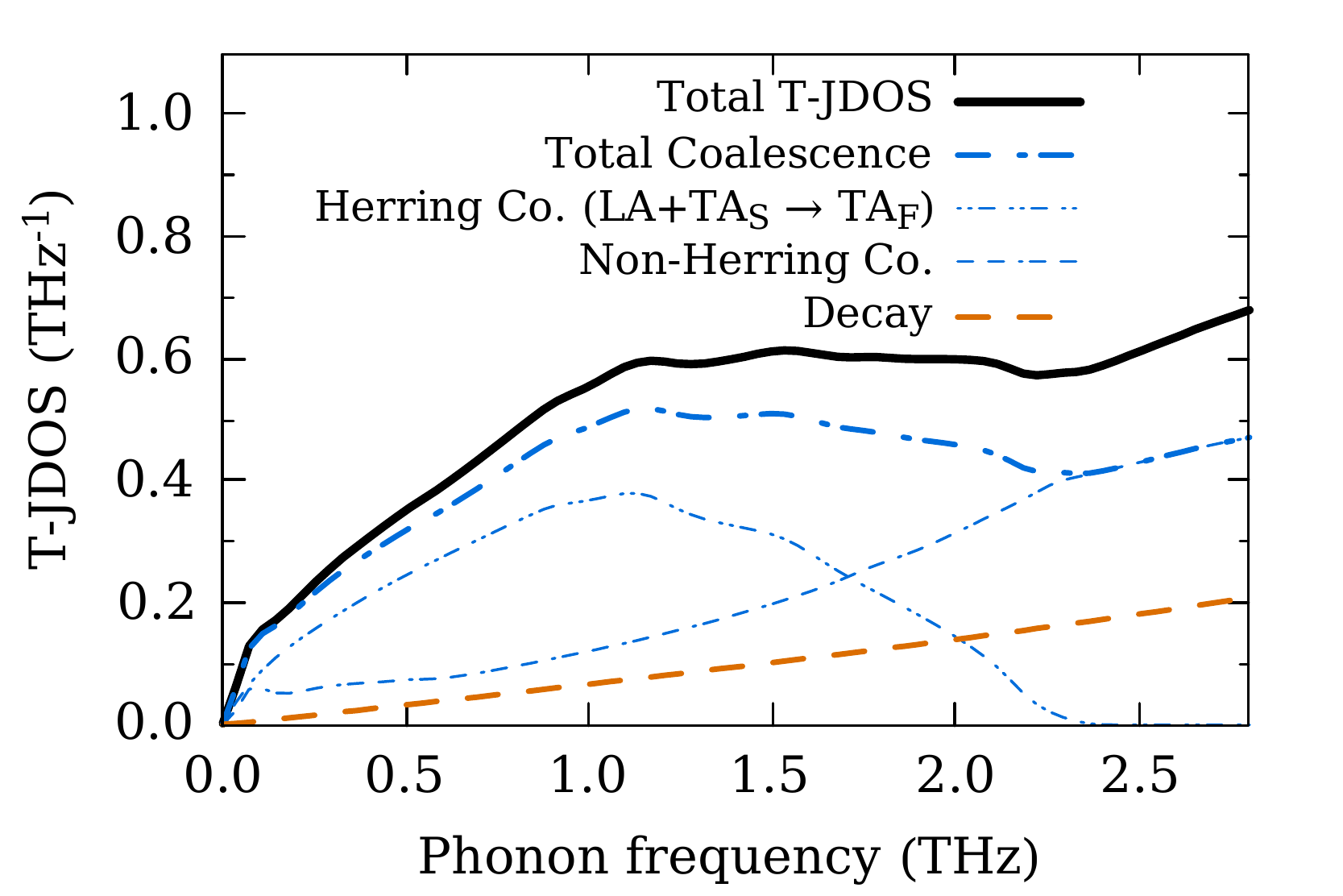}}
\caption{Local joint phonon density of states (L-JDOS) for Silicon at 50\,K and 300\,K along directions {$[100]$} and {$[111]$}. Same comment as figure 3.}
\end{figure}

\bibliography{bibliography}